\documentclass[a4paper,fleqn,usenatbib]{mnras}

\usepackage[T1]{fontenc}
\usepackage{ae,aecompl}

\usepackage{natbib,twoopt}
\usepackage[fleqn]{amsmath}
\usepackage{graphicx}
\usepackage{xcolor}
\usepackage{scalefnt}

\renewcommand{\vec}[1]{\textbf{#1}}
\title[Growth of satellites in a Jovian massive disc]{Growth and evolution of satellites in a Jovian massive disc}

\author[R. A. Moraes, W. Kley and E. Vieira Neto]{%
  R. A. Moraes$^{1,2}$\thanks{E-mail: ricardo.moraes07@gmail.com (RAM)},
  W. Kley$^{2}$ and
  E. Vieira Neto$^{1}$\\
  $^{1}$ UNESP, Univ. Estadual Paulista - Grupo de Din\^{a}mica Orbital \& Planetologia, Guaratinguet\'{a}, CEP 12.516-410, S\~{a}o Paulo, Brazil\\
  $^{2}$ Institut f\"{u}r Astronomie und Astrophysik, Universit\"{a}t T\"{u}bingen, Auf der Morgenstelle 10, D-72076 T\"{u}bingen, Germany}
\date{Accepted XXX. Received YYY; in original form ZZZ}

\pubyear{2017}

\begin{document}
\label{firstpage}
\pagerange{\pageref{firstpage}--\pageref{lastpage}}
\maketitle


\begin{abstract}
The formation of satellite systems in circum-planetary discs is considered to be similar to the formation of rocky planets in a proto-planetary disc, especially Super-Earths. Thus, it is possible to use systems with large satellites to test formation theories that are also applicable to extrasolar planets. Furthermore, a better understanding of the origin of satellites might yield important information about the environment near the growing planet during the last stages of planet formation. In this work we investigate the formation and migration of the Jovian satellites through N-body simulations. We simulated a massive, static, low viscosity, circum-planetary disc in agreement with the minimum mass sub-nebula model prescriptions for its total mass. In hydrodynamic simulations we found no signs of gaps, therefore type II migration is not expected. Hence, we used analytic prescriptions for type I migration, eccentricity and inclination damping, and performed N-body simulations with damping forces added. Detailed parameter studies showed that the number of final satellites is strong influenced by the initial distribution of embryos, the disc temperature, and the initial gas density profile. For steeper initial density profiles it is possible to form systems with multiple satellites in resonance while a flatter profile favours the formation of satellites close to the region of the Galilean satellites. We show that the formation of massive satellites such as Ganymede and Callisto can be achieved for hotter discs with an aspect ratio of $H/r \sim 0.15$  for which the ice line was located around $30$ $R_J$.
%%% and the migration rate was moderated.
\end{abstract}
%%%%%%%%%%%%%%%%%%%%%%%%%%%%%%%%%%%%%%%%%%%%%%%%%%

%%%%%%%%%%%%%%%%% BODY OF PAPER %%%%%%%%%%%%%%%%%%

\section{Introduction}
\label{sone}
Orbiting Jupiter we have the most famous family of regular satellites in our Solar System, the Galilean satellites. These bodies are in almost circular and coplanar orbits, rotating in the same direction as the rotation movement of Jupiter. Such characteristics led to a well accepted theory that these bodies were formed in a circum-planetary disc around Jupiter \citep{Lunine-Stevenson-1982}. This formation scenario somehow resembles the scenario proposed for formation of our Solar System, for such that the Galilean system is often called a ``mini Solar System". Studying the formation of this satellite system could provide information about the last stages of formation of Jupiter. Nowadays there are two competitive models that aim to describe the origin of the regular satellites in a circum-planetary disc, the minimum mass sub-nebula model (MMSN) \citep{Lunine-Stevenson-1982, Mosqueira-Estrada-2003a, Mosqueira-Estrada-2003b, Estrada-etal-2009} and the gas-starved disc model \citep{Canup-Ward-2002, Canup-Ward-2006, Canup-Ward-2009}.\par
The gas-starved disc model proposes an initially low mass disc that is continuously supplied by the infall of material from the circum-stellar proto-planetary disc. In this model the formation of the satellites takes place at the same time as the formation of the parent planet. \citet{kley-1999} and \citet{lubow} showed that after the forming planet opens its gap in the disc, the flow of mass toward the planet is not halted and the planet continues to grow from this material. This continuous inflow of material will provide a source for the satellites to grow, always maintaining a low surface density disc with a peak of 100 g/cm$^2$. In this model, the first generation of satellites forms and performs a type I migration through the disc, then a second generation of satellites forms and repeat the same process. In this way the Galilean satellites were only the last generation of satellites formed in the circum-planetary disc around Jupiter. The low surface density increases the time of formation of the Galilean satellites, using the process of inflow of material to add the amount of mass needed to form massive bodies, this feature corroborates the idea of a late formation of Callisto (the outermost satellite). However \citet{Mosqueira-Estrada-2003b} argued that this mechanism could not only delay the process of satellite formation but prevent it. The gas-starved disc model is widely used by several authors to study the formation of regular satellites in other planets of our Solar System and in extrasolar systems \citep{Heller-Pudritz-2015a, Heller-Pudritz-2015b}.\par
A competitive model for the gas-starved disc model is the minimum mass sub-nebula model. This model was first proposed to describe the formation of the planets system in our Solar System \citep{Weidenschilling-1977, Hayashi-1981} and was later adapted for satellite systems. In the MMSN model, a massive disc with a peak surface density around $10^5$ g/cm$^2$ is accumulated by infall of gas through the Lagrangian points of the planet at the late stages of its formation. In this case, the disc is static, with no inflow of material and has a low turbulence during the evolution of the satellites, in \citet{Mosqueira-Estrada-2003a}, for example, the authors considered an isolated disc where the turbulence does not play a role any more. 
Up to date, there is no consensus about which model can describe better the formation of satellites and several semi-analytical studies had been performed in order to explore more parameters of both models \citep{Sasaki-etal-2010, Ogihara-Ida-2012, Miguel-Ida-2016}.\par
The properties and characteristics of the Galileans satellites are very peculiar. Starting with their sizes and masses, these satellites are among the most massive in our Solar System (Table \ref{tab:properties}) including Titan and Triton, orbiting Saturn and Neptune, respectively.
The largest satellite, Ganymede, is more massive than the planet Mercury. The composition of these satellites and their fraction of ice is different, which means that the temperature in the disc during the formation of the satellites was different. The increasing ice fraction per distance in the Galilean system may imply that near the planet the temperature was higher than farther out. \par   
\begin{table}
\caption[Properties]{Properties and fraction of rock and ice of the Galileans satellites, where $a$, $M$ and $e$ are the semi-major axis (in units of the radii of Jupiter), the mass (in units of $10^{-5}$ masses of Jupiter) and the eccentricity of the satellites, respectively.}
\begin{tabular}{c|c|c|c|c|c}\hline 
Satellites  & $a$					& $M$ 						   & $e$      & Rock  		& Ice			\\
			& $R_{J}$				&$10^{-5}M_{J}$ 			   & 		  &	$\%$	  	& $\%$			\\ \hline	
Io          &  5.9                  & 4.7                          & 0.0041   &  100        & 0         \\ 
Europa      &  9.4                  & 2.5                          & 0.01     &  $\sim$ 82  & $\sim$ 8  	\\ 
Ganymede    &  15.0                 & 7.8                          & 0.0015   &  $\sim$ 55  & $\sim$ 45 	\\ 
Callisto    &  26.4                 & 5.7                          & 0.007    &  $\sim$ 44  & $\sim$ 56 	\\ \hline
\end{tabular}
\label{tab:properties}
\end{table}
The current dynamical configuration of the Galileans is also interesting. Io, Europa and Ganymede are locked in a 1:2:4 mean motion resonance. In this way the successive conjunction of these bodies occurs always close to the same longitude. The Galilean satellites are also in the Laplace resonance with resonant angle defined as,  
\begin{align}\label{eq:laplace}
\phi = \lambda_I-3\lambda_E+2\lambda_G
\end{align}
where $\lambda_I$, $\lambda_E$ and $\lambda_G$ are the mean longitude of Io, Europa and Ganymede, respectively. In the case of these satellites, the resonant angle librates around $180^{\circ}$ ($\phi\approx180^{\circ}$). This implies that when Europa and Ganymede are in conjunction (with respect to Jupiter), i.e., $\lambda_E=\lambda_G$, Io will be in the opposite position, because $\lambda_I-\lambda_E=\lambda_I-\lambda_G\approx 180^{\circ}$. In this way a triple conjunction is impossible.\par
The resonances in the Galilean system is a signature of orbital migration during or after their formation \citep{Ogihara-Ida-2012}. In order to explain the resonances using orbital migration two different theoretical scenarios were proposed. In the first scenario, \citet{Yoder-1979} and \citet{Yoder-Peale-1981} argued that Io migrates outward due to the gravitational tides from Jupiter and eventually captured Europa in a 2:1 mean motion resonance, after this point the orbital expansion continues and Europa encountered Ganymede, forming a new 2:1 commensurability, finally resulting in the Laplace resonance. In the second scenario, \citet{Greenberg-1987} and \citet{Peale-Lee-2002} associated the resonances to a inward type I migration, in this case the most massive satellite, Ganymede, performed a rapid type I migration through the disc until capture the inner satellites in the Laplace resonance. The second scenario as it was initially proposed has problems, \citet{Canup-Ward-2006} showed through N-body simulations, that when a satellite in the outer disc reaches a mass comparable with Ganymede, the satellites in the inner disc are less massive than Io and Europa. In this way, a possible solution could be the migration of a Ganymede-size satellite after the satellites in the inner region have been already formed.\par
In this work we study the scenario of formation of satellite systems with the characteristics of the Galileans under the assumptions of the classical MMSN model. We performed several N-body simulations using the package Mercury \citep{Chambers-1999} including type I migration and eccentricity and inclination damping adapted from \citet{Cresswell-Nelson-2008} to act on the proto-satellites embryos, and an aero-dynamical drag \citep{Adachi-etal-1976} which will act over the satellitesimals. Throughout the simulations we explored different parameters such as the radial distribution and masses of the solids, density profile of the gaseous nebula and the aspect ratio of the disc. \par
This paper is organized as follows, in section \ref{stwo} we explain the model and show the parameters used. The results of our simulations are presented in section \ref{sthree} and in section \ref{conclusion} we point out our conclusions.

\section{The Model}\label{stwo}
According to the MMSN model the Galilean satellites will form in a circum-planetary disc around Jupiter, during the late stages of formation of the planet. At this point the gas infall onto Jupiter is already finished and the turbulence in the circum-planetary disc is weak. We will apply these premises to our model as will be described below.\\
\subsection{System}
We considerer a coordinate system centred on the planet without the effects of the Sun. As we are working at the final stages of Jupiter's formation, the circum-planetary disc is immersed in the gap opened by the planet. Unless we state differently, the circum-planetary disc will be composed of gas, satellitesimals and satellite embryos. Following the MMSN assumptions, the total mass of gas and solids (satellitesimals and embryos) in the disc will be $M_{gas} \sim 2\times10^{-2}$ $M_{Jup}$ and $M_{solid} \sim 2\times10^{-4}$ $M_{Jup}$, respectively. All the solids in the disc are affected by the presence of the gas, the embryos will gravitationally interact with each other and with the satellitesimals while the gravitational interaction between satellitesimals will be neglected.\\
Because of the uncertainties in the initial solid surface density we decided to neglect the increase of solid surface density due to ice condensation \citep{Miguel-Ida-2016}, also we choose the gas to dust ratio to be equal to 100, analogous to the solar nebula\footnote{Other gas to dust ratios could be tested, usually keeping the total mass of solids in the disc and decreasing the gas surface density. These models are known as solid-enhanced minimum-mass model \citep{Mosqueira-Estrada-2003b, Estrada-etal-2009}.}.
In section~\ref{subsec:enhanced} we discuss a model with enhanced amount of solids.
The density profile and the radial distribution will be addressed later, once the characteristics of the solids disc will be different for each simulation.
\subsection{The Disc Structure}  %%WK: new title
As pointed out by \citet{Miguel-Ida-2016} there are many factors that should be taken into account when we define the gas surface disc distribution for a circum-planetary disc around Jupiter, such as the infall and viscous accretion onto the planet. As we are not interested in developing a very sophisticated model for the Jupiter's nebula, we will argue that when a satellitesimal (or embryo) becomes large enough to decouple from the inflowing gas, the drag forces generated by the gas in the disc will not play an important role when these bodies cross the denser parts of the disc, which would be in the vicinity of the planet. Hydrodynamic simulations had shown that for these locations the gas distribution is approximately axisymmetric \citep{Tanigawa-etal-2012}. Thus we assume the following radial vertical distribution for the gas density of the circum-planetary disc,
\begin{align}\label{eq:dens1}
\rho_{gas}(r,z) = \dfrac{\Sigma_{gas}(r,t)}{\sqrt{\pi} H}\exp \left(\dfrac{-z^2}{H^2}\right)
\end{align}
where $\Sigma_{gas} (r,t)$ and $H$ are the gas surface density and the scale height of the circum-planetary disc, respectively, given by
\begin{align}\label{eq:dens2}
\Sigma_{gas}(r,t) = \Sigma_{gas,t} \left(\dfrac{r}{R_0}\right)^{-\beta}
\end{align}
and 
\begin{align}\label{scale1}
H = 0.05 r \,.
\end{align}
The length scale, $R_0=150 R_J$, corresponding to the effective size of the circum-planetary disc, 
$\Sigma_{gas,t}$ is given by Eq. \ref{eq:dens3} and $\beta = 3/2$ is the local surface density profile.\\
For our disc model we adopted an exponential prescription in time for the gaseous disc dissipation given by \citep{Ida-Lin-2004}
\begin{align}\label{eq:dens3}
\Sigma_{gas,t} = \Sigma_{gas,0} \exp\left(-\dfrac{t}{\tau_{disc}}\right),
\end{align}
where $t$ is time, $\Sigma_{gas,0}$ is the initial density at $R_0$ and $\tau_{disc}$ is the global disc diffusion time-scale given by
\begin{align}\label{eq:tau}
\tau_{disc} \sim 3\times 10^{5}\dfrac{R_{disc}}{100R_{J}} \left(\dfrac{\alpha}{10^{-5}}\right)^{-1} yrs.
\end{align} 
The viscous parameter $\alpha$ depends of the turbulence in the disc and as shown by \citet{Turner-etal-2014}, \citet{Fujii-etal-2014} and, more recently, \citet{Fujii-etal-2017}, there is an absence of magneto-rotational instability (MRI) active regions in the circum-planetary disc, which suggest a value for $\alpha$ smaller than $10^{-3}$. Here, we adopted $\alpha$ between $10^{-4}$ and $10^{-5}$
\citep{Stoll2014A&A...572A..77S}, which implies $\tau_{disc}$ proportional to values between  $10^{4}$ to $10^{5}$ years.

In our disc models we use a given temperature profile for the disc, for which we assume a constant aspect ratio.
In the first set of models we use $h=0.05$ (a value often used for the Solar nebula) which results in a 
temperature profile for the disc given by
\begin{align}\label{eq:temp1}
T = T_0\left(\dfrac{r}{R_J}\right)^{-1},
\end{align}  
where $T_0 = 545$ K is the temperature at $1$ $R_J$. In a second set of models we use a higher aspect ratio, $h=0.15$, which places the iceline at about $30 R_J$, in this case $T_0\sim 4902$ K.
\subsection{Equations of Motion}
We performed N-body simulations for our numerical calculations with the package Mercury \citep{Chambers-1999}. In all the simulations, when two bodies collide we consider perfect merging with conservation of angular momentum. \\
In planetocentric coordinates the equations of motion of the particle $k$ at distance $r_k$ are,
\begin{align}\label{eq:motion1}
\dfrac{d^2\vec{r}_k}{dt^2} = -GM_p\dfrac{\vec{r}_k}{|\vec{r}_k|^3} - \sum\limits_{i\neq k}GM_i\dfrac{\vec{r}_k-\vec{r}_i}{|\vec{r}_k-\vec{r}_i|^3} - \sum\limits_i GM_i\dfrac{\vec{r}_i}{|\vec{r}_i|^3}\nonumber \\
+ \vec{F}_{mig} + \vec{F}_{damp,e} + \vec{F}_{damp,I} + \vec{F}_{drag}
\end{align}
where $k, i = 1, 2, \ldots, n$. The terms on the right side of Eq. \ref{eq:motion1} are the gravitational force from the planet (central body), the mutual gravity interaction between the bodies, the indirect terms, the tidal damping in the semi-major axis for the satellites embryos (type I migration), eccentricity damping, inclination damping and aero-dynamic gas drag over satellitesimals, respectively.
\subsection{Gas Effects}
Satellitesimals and embryos interact with the circum-planetary disc differently. The satellitesimals are small and feel only the headwind from the gas, the embryos are massive enough to produce density waves in the disc, and are affected by torques from these waves. We will use different approaches to take into account the gas interaction with the different bodies.\\ 
The aero-dynamical drag is responsible for transfer angular momentum from the gas to the satellitesimals. Generally the gas is rotating with a sub-keplerian speed and the interaction with the satellitesimals causes these bodies to migrate inward. In our models we implemented the aero-dynamical gas drag acceleration given by \citet{Adachi-etal-1976},
\begin{align}\label{eq:adachi1}
\vec{a}_{drag} = -\dfrac{3C_d\rho_{gas}v_{rel}}{8\rho_{sat}R_{sat}}\vec{v}_{rel} 
\end{align}
with $C_d$ being the drag coefficient calculated as described by \citet{Brasser-etal-2007}, $v_{rel}$ is velocity of the satellitesimals with respect to the gas, $\rho_{sat}$ and $R_{sat}$ are the density and the radius of the bodies. The intensity of the drag dramatically depends of the radius of the bodies, in our simulations we set the radius of all satellitesimals to 10 km.\\   
To reproduce the effects of the gaseous disc on the satellite embryos we will separate the effects in two categories, considering that the eccentricity and inclinations of the embryos can be different than zero. First, the response of the embryos to the density waves produced by the embryos cause those bodies to exchange angular momentum with the disc and migrate inward in a type I regime. Second, if the embryos are eccentric and/or inclined the interaction with the disc will damp both eccentricity and inclination leading to circular and coplanar embryos \citep{Izidoro-etal-2014}.\\
To properly take those effects into account we will follow the formalism of \citet{Tanaka-Ward-2004}, generalized by \citet{Papaloizou-Larwood-2000} and \citet{Cresswell-Nelson-2008} to include high eccentricities, defining a characteristic time-scale,
\begin{align}\label{eq:twave1}
t_{wave} = \dfrac{M_p}{M_{emb}}\dfrac{M_p}{\Sigma_{gas}a^2}h^4\Omega^{-1}_{k},
\end{align} 
$M_p$ is the mass of the planet, $h=H/r=0.05$ is the aspect ratio of the disc and $M_{emb}$, $a$ and $\Omega_{k}$ are the mass, the semi-major axis and the keplerian frequency of the satellite embryos. \\
For the time-scales for the migration, eccentricity and inclination damping we used the equations from \citet{Izidoro-etal-2016}, derived from the numerical fits from hydrodynamic simulations found by \citet{ Cresswell-Nelson-2008}, and given by
\begin{align}\label{eq:tmig1}
t_{mig} = \dfrac{2.0}{2.7+1.1\beta}\left(\dfrac{1.0+\left(\dfrac{e}{1.3h}\right)^5}{1.0-\left(\dfrac{e}{1.1h}\right)^4}\right)t_{wave}h^{-2},
\end{align} 
\begin{align}\label{eq:te1}
t_{e} = \dfrac{t_{wave}}{0.780}\left(1.0-0.14\left(\dfrac{e}{h}\right)^2+0.06\left(\dfrac{e}{h}\right)^3+0.18\left(\dfrac{e}{h}\right)\left(\dfrac{I}{h}\right)^2\right)
\end{align}
and
\begin{align}\label{eq:ti1}
t_{I} = \dfrac{t_{wave}}{0.544}\left(1.0-0.30\left(\dfrac{I}{h}\right)^2+0.24\left(\dfrac{e}{h}\right)^3+0.14\left(\dfrac{e}{h}\right)^2\left(\dfrac{I}{h}\right)\right)
\end{align}
with $e$ and $I$ being the embryos eccentricity and inclination while $\beta$ is the local surface density profile.\\
From Eqs. \ref{eq:tmig1}, \ref{eq:te1} and \ref{eq:ti1} we can write the synthetic accelerations exerted by the disc onto the planet \citep{Cresswell-Nelson-2008} as
\begin{align}\label{eq:am1}
\vec{a}_{mig} = -\dfrac{\vec{v}}{t_{mig}},
\end{align} 
\begin{align}\label{eq:ae1}
\vec{a}_{damp,e} = -2\dfrac{\left(\vec{v}\cdot\vec{r}\right)}{r^2t_{e}}\vec{r},
\end{align} 
and
\begin{align}\label{eq:ai1}
\vec{a}_{damp,I} = -\dfrac{v_z}{t_{i}}\vec{k}
\end{align} 
where $\vec{k}$ is the unit vector in the z-direction.\\
The accelerations described in Eqs. \ref{eq:adachi1}, \ref{eq:am1}, \ref{eq:ae1} and \ref{eq:ai1} can easily be translated into the forces presented in the Eq. \ref{eq:motion1}.\\
The interaction of the planetary magnetic field with the disc was studied by \citet{Takata-Stevenson-1996} and the authors found that the circum-planetary disc is truncated at the co-rotation radius of the planet. Thus, following the prescription of \citet{Sasaki-etal-2010} we introduced an inner cavity in the disc at $\sim 2.25$ $R_J$, inside this cavity the effects of the gas drag will be neglected.
\subsection{Hydrodynamic Simulations}   %% WK: Type II Migration
In the present work we will neglect type II migration because, in our case, the embryos will not become massive enough to open a gap in the disc. The idea that the type II migration could be responsible for the evolution of a satellite system as the Galileans was explored by \citet{Miguel-Ida-2016}. They argued that in a low viscous disc a gap will be easily opened by the satellites at relatively small masses. Using the results that indicate absence of MRI-active regions in the disc \citep{Turner-etal-2014,Fujii-etal-2014,Fujii-etal-2017}, \citet{Miguel-Ida-2016} choose an $\alpha$ value between $10^{-4}$ and $10^{-6}$, which we considered excessive small.\\
To verify the presence of a gap in a low viscous disc we performed some hydrodynamic simulations to study the evolution of the disc in the presence of a group of satellites with the same mass as the Galileans.\\
For our hydrodynamic simulations we used the FARGO 2D package \citep{masset} with a planet of one Jupiter mass in the centre of the system. We simulate an locally isothermal disc around Jupiter with scale height $h=0.05$, the disc extends from $r_{in} = 2.25$ $R_J$ to $r_{out} = 150$ $R_J$. The surface density profile adopted was $\Sigma = \Sigma_0r^{-3/2}$, with the initial surface density $\Sigma_0 = 1.48\times 10^{-4}$ $M_J/R_J^2$, such as the disc contains $2\times10^{-2}$ $M_J$. As the circum-planetary disc must have low viscosity we used the $\alpha$-prescription of \citet{shakura} with $\alpha = 5\times 10^{-4}$, a value derived for purely hydrodynamic viscosity by the viscous shear instability, \citep{Stoll2014A&A...572A..77S, Stoll-etal-2017a, Stoll-etal-2017b}. For the grid resolution we adopted $(N_r, N_{\phi}) = (300, 300)$.\\
The four satellites embedded in the disc were initialized with the current mass of the Galileans (see Table \ref{tab:properties}), in circular orbits and far from their current position, in this way they can migrate through the disc. The accretion of gas by the satellites was neglected because it is believed that these satellites are not composed by gas. The simulation was performed for 150 years.
\begin{figure*}
  \begin{center}
  \includegraphics[width=\textwidth]{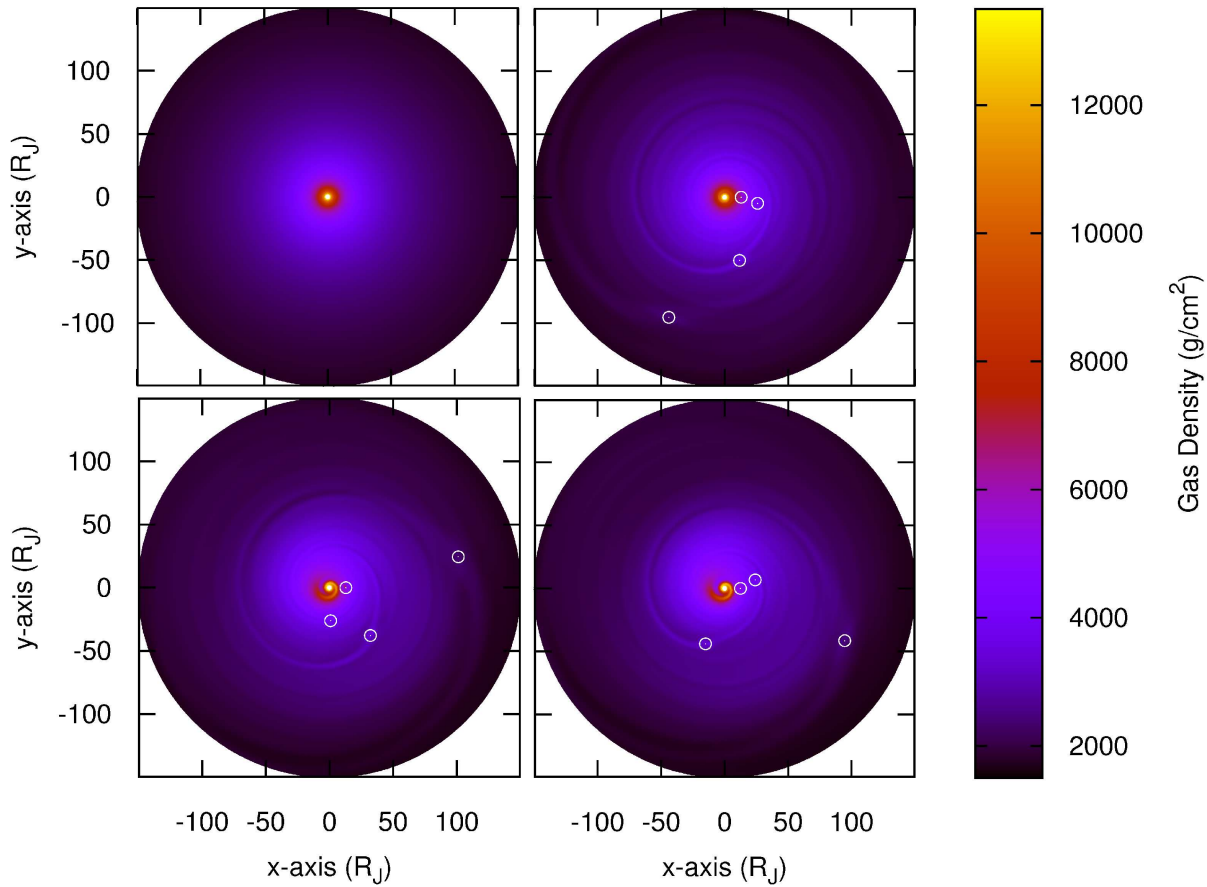}
  \caption[Density.]{Surface density of the disc around Jupiter, after 0, 50, 100 and 150 years from upper left side to down right side, respectively. The white open circles indicate the position of the satellites.}
  \label{fig:density} 
  \end{center} 
\end{figure*}
In Fig. \ref{fig:density} we present snapshots for the evolution of the system for 0, 50, 100 and 150 years. The structure of the disc is not changed drastically by the presence of the satellites, it is possible to identify some spiral arms near the satellites. These structures are the response of the disc to the presence of bodies, and due to the exchange of angular momentum between the bodies and the disc the satellites migrated inward. \par
Our results did not show any sign of gap formation around the satellites. Thus, there is no apparent reason to use the type II regime to describe the migration of the satellites. None the less, we highlight that, due to computational limitation, our simulation only covered the early phases of interaction between the satellites and the disc. A disc with lower viscosity and/or massive satellites might produce different results, but we believe that these conditions are very unrealistic and might lead to an equivocated understanding of the problem. \par

\section{Results}\label{sthree}
In this section we will present our models and follow the growth and orbital evolution of satellite embryos using different set-ups. 
In the following, we will explore the impact of some parameters as distribution of the embryos, mass and number of the satellitesimals, the density profile of the gas disc and the aspect ratio of the disc. A summary of our models is shown in Table \ref{tab:simulations}.
\begin{table*}
\caption[Models]{Parameters of the simulations. Where Emb. is the number of embryos, Satel. is the number of satellitesimals, $M_{emb}$, $I_{emb}$, $e_{emb}$, $a_{emb}$  are the initial mass, inclination, eccentricity and semi-major axis of the embryos,  $M_{satel}$, $a_{satel}$ are the mass and the semi-major axis of the satellitesimals, $\beta$ is the local surface density profile, $\dfrac{M_{gas}}{M_{solid}}$ is the gas to dust ratio and $h$ is the aspect ratio of the disc.}
\begin{tabular}{c|c|c|c|c|c|c|c|c|c|c|c}\hline 
Models	    	 		& Emb.    & Satel.	& $M_{emb}$ 	& $M_{satel}$  		 & $I_{emb}$ & $e_{emb}$ & $\beta$ & $a_{emb}$ 		& $a_{satel}$ 	& $\dfrac{M_{gas}}{M_{solid}}$ & $h$	\\
						& 	  		 &				   	& $10^{-5} M_J$	& $10^{-5} M_J$ 	 & Deg	 	 & 	  	 	 &		   & $R_J$ 			&$R_J$ 			&				&			\\ \hline	
\texttt{4-sat-1}        &  4         & 0               	& 2.5 - 7.8     &   -  				 & 0.0		 & 0.0		 & 1.5     &13.4 - 107.2 	& -				& 58			& 0.05			\\ 
\texttt{4-sat-2}      	&  4         & 0               	& 2.5 - 7.8     &   -    			 & 5.0		 & 0.1		 & 1.5     &13.4 - 107.2 	& -				& 58			& 0.05			\\ 
\texttt{multi-sat-1}    &  20        & 2000            	& 0.04 - 0.17   &  0.01    			 & 0.0		 & 0.0		 & 1.5     &21.0 - 170.0 	& 21.0 - 170.0 	& 100 			& 0.05			\\ 
\texttt{multi-sat-2}    &  20        & 2500            	& 0.04 - 0.17   &  0.007  			 & 0.0		 & 0.0		 & 1.5     &21.0 - 170.0 	& 21.0 - 170.0 	& 100			& 0.05			\\ 
\texttt{multi-sat-3}    &  20        & 2000            	& 0.1 - 0.43    &  0.007			 & 0.0		 & 0.0		 & 1.5     &21.0 - 150.0 	& 2.25 - 150.0 	& 100			& 0.05			\\ 
\texttt{multi-sat-uni-1}&  20        & 2000            	& 0.04 - 0.17   &  0.01    			 & 0.0		 & 0.0		 & 0.0     &21.0 - 170.0 	& 21.0 - 170.0 	& 100			& 0.05			\\
\texttt{multi-sat-uni-1-h015}&  20        & 2000            	& 0.04 - 0.17   &  0.01    			 & 0.0		 & 0.0		 & 0.0     &21.0 - 170.0 	& 21.0 - 170.0 	& 100	& 0.15					\\
\texttt{multi-sat-uni-2-h015}&  20        & 2500            	& 0.04 - 0.17   &  0.007  			 & 0.0		 & 0.0		 & 0.0     &21.0 - 170.0 	& 21.0 - 170.0 	& 100	& 0.15				\\  
\texttt{multi-sat-uni-3-h015}&  20        & 2000            	& 0.1 - 0.43    &  0.007			 & 0.0		 & 0.0		 & 0.0     &21.0 - 150.0 	& 2.25 - 150.0 	& 100	& 0.15				\\  
\texttt{multi-sat-uni-4-h015}&  20        & 2000            	& 0.04 - 0.17   &  0.01    			 & 0.0		 & 0.0		 & 0.0     &21.0 - 170.0 	& 21.0 - 170.0 	& 10	& 0.15					\\ 
\texttt{multi-sat-uni-5-h015}&  20        & 2000            	& 0.02 - 0.28   &  0.01    			 & 0.0		 & 0.0		 & 0.0     &21.0 - 170.0 	& 21.0 - 170.0 	& 100	& 0.15					\\ \hline
\end{tabular}
\label{tab:simulations}
\end{table*}
The first two models in Table~\ref{tab:simulations} (\texttt{4-sat-1} and \texttt{4-sat-2}) 
describe evolutions of 4 embryos with the present mass of the Galilean satellites, 
i.e. no mass growth of the embryos by satellitesimals was allowed.
The next 3 models (\texttt{multi-sat-1}, \texttt{multi-sat-2}, \texttt{multi-sat-3}) start with initially 20 embryos having a small fraction of the mass of the
Galilean moons and
2000 to 2500 satellitesimals that were initially spread throughout the whole disc. In the model \texttt{multi-sat-uni-1} we simulate the evolution of the system in an uniform gas density nebula ($\beta=0$) using the same set-up of model \texttt{multi-sat-1}.
In the models \texttt{multi-sat-uni-1-h015}, \texttt{multi-sat-uni-2-h015} and \texttt{multi-sat-uni-3-h015} we ran the same set-ups of models \texttt{multi-sat-1}, \texttt{multi-sat-2} and \texttt{multi-sat-3}, respectively, using an uniform
gas density with $\beta = 0$ and a higher aspect ratio, $h=0.15$.
In the model (\texttt{multi-sat-uni-4-h015}) we simulate the evolution of satellites in a solids-enhanced disc with uniform gas density and in the model \texttt{multi-sat-uni-5-h015} was performed a simulation with uniform gas density and steep distribution of embryos, for both models $h$ was set to $h=0.15$.
%
%%%%%%%%%%%%%%%%%%%%%%%%%%%%%%%%%%%%%%%%%%%%%%%%%%%%%%%%%%%%%%%%%%%%%%%%%%%%%%%%%%%%%%%%%%%%%%%%%%%%%%%%%%%%%%%%
\subsection{Migration of Four Satellites (Model \texttt{4-sat-1})}
\begin{figure} 
  \begin{center}
  \includegraphics[scale=0.67]{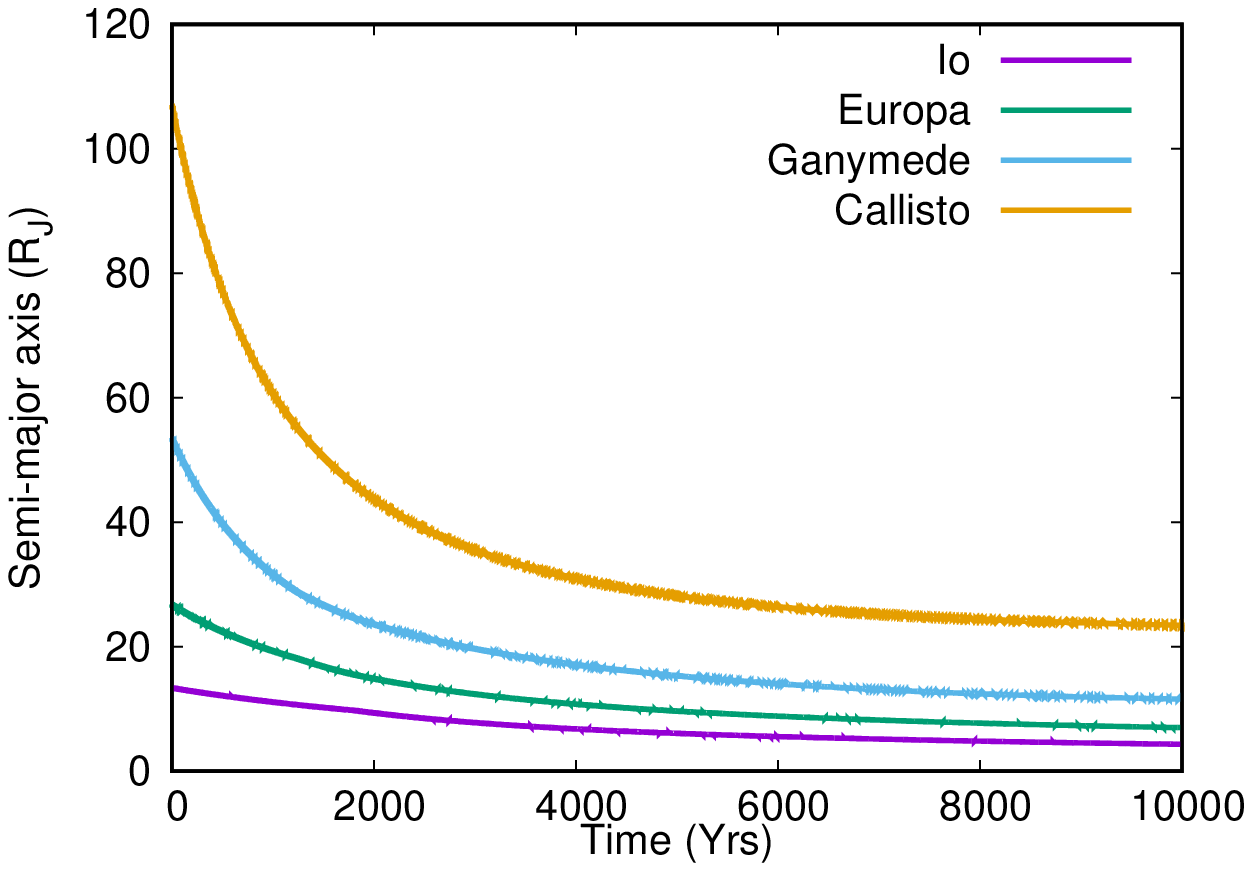}
  \includegraphics[scale=0.67]{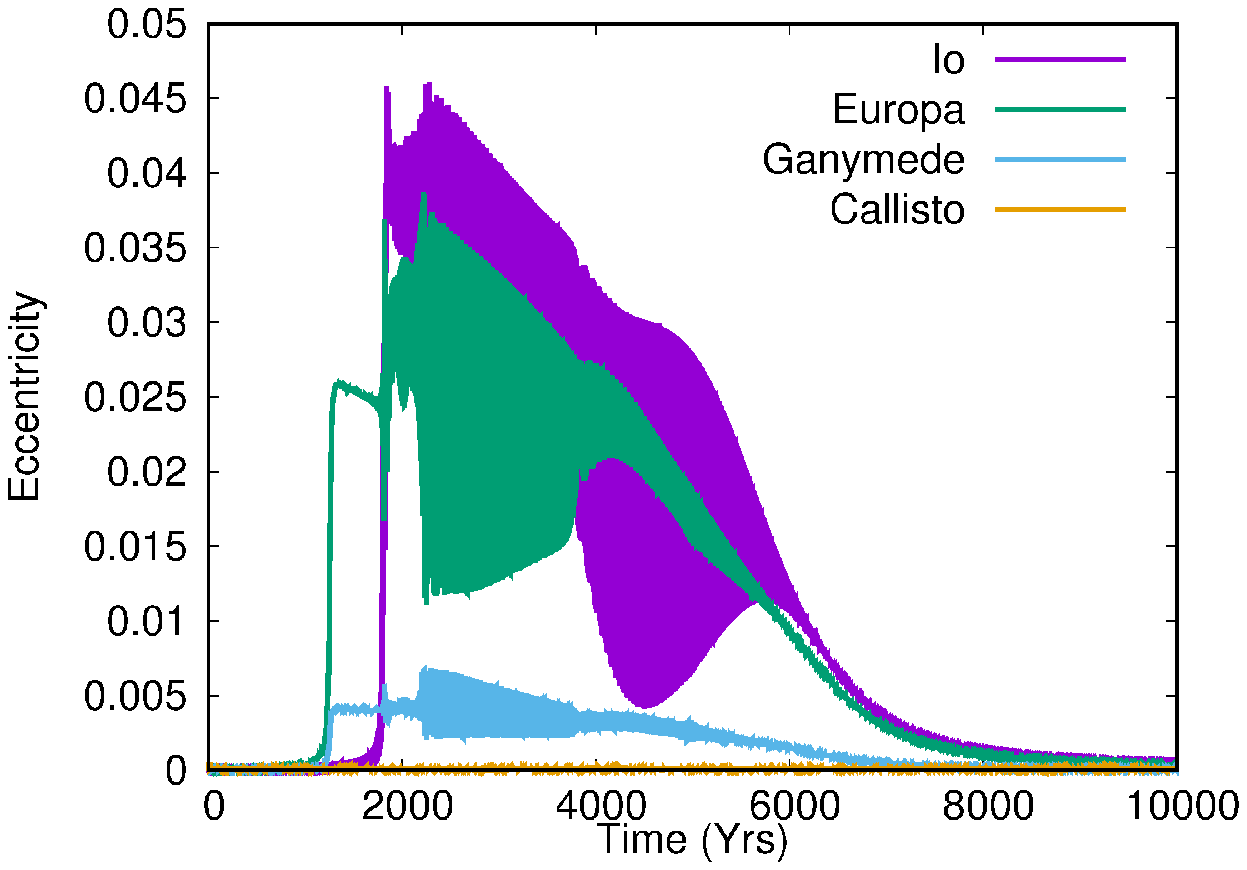}
  \caption[Semi-ecc.]{Evolution of the semi-major axis and eccentricity of the satellites for model \texttt{4-sat-1} over 10000~yrs. }
  \label{fig:semi-ecc-model-52} 
  \end{center} 
\end{figure}

In this model we will simulate only the migration of a set of massive satellites having the same mass as the Galileans satellites today. The surface density of the gaseous disc was described in section \ref{stwo}. Once we are simulating a phase when the satellites are already formed it is expected that these bodies performed a previous migration and a portion of the gaseous disc was also dissipated. In this way, we chose the mass of the gaseous disc to be 60\% of the total mass proposed by the MMSN model. \\
Currently, the three inner Galilean satellites are in a 1:2:4 mean motion resonance. This implies a radial separation of a factor of $\sim 1.6$ from one satellite to another. In our model the four satellites were initially separated by a factor of 2 from $13.4$ $R_J$ to $107.2$ $R_J$, in circular and coplanar orbits. During the evolution of the satellites we will check if they can reach the resonant configuration naturally. The results for the semi-major axis and eccentricity evolution are presented in Fig. \ref{fig:semi-ecc-model-52}. \par
The damping in the eccentricity kept the satellites almost circular during the evolution of the system, however it did not prevent spikes in the eccentricity of Io, Europa and Ganymede, when these bodies reach a resonant configuration. \par
Due to the interaction with the gaseous nebula the satellites performed an inward migration, the final positions of the satellites from the innermost to the outermost are $4.31$ $R_J$, $6.85$ $R_J$, $10.9$ $R_J$ and $23.3$ $R_J$, respectively. Once small satellitesimals were not considered, the outer satellites migrated freely through the disc until approach the two innermost ones. At this point, the migration rate of these bodies decreases due to gravitational interaction with the inner satellites and they slowly end in a resonant chain. At the end of the simulation the three innermost satellites are locked in a 1:2:4 mean motion resonance. As shown by \citet{Ogihara-Ida-2012} at late stages of satellite evolution the gas surface density would be less denser, such as the damping on eccentricity and semi-major axis will enable the satellites to get lock in multiples 2:1 resonances if the satellites are massive, which is the case here. However, the same authors found that different configurations, with more formed satellites, could yield satellites in different first order resonant configurations, such as 3:2 and 4:3 for example. We believe that if more than four formed satellites were simulated under the same conditions we used, multiple resonant chains would be found and some of these resonances could be different from 2:1.

Our first scenario is very specific because we assume that the satellites were already formed in circular and coplanar orbits, and the satellitesimal disc was not considered. However, we verify that under such assumptions type I migration is enough to drive the satellites from outer locations to the position of the Jovian satellites.
%%%%%%%%%%%%%%%%%%%%%%%%%%%%%%%%%%%%%%%%%%%%%%%%%%%%%%%%%%%%%%%%%%%%%%%%%%%%%%%%%%%%%%%%%%%%%%%%%%%%%%%%%%%%%%%%
\subsection{Migration of Four Eccentric and Inclined Satellites (Model \texttt{4-sat-2})}
\begin{figure} 
  \begin{center}
  \includegraphics[scale=0.67]{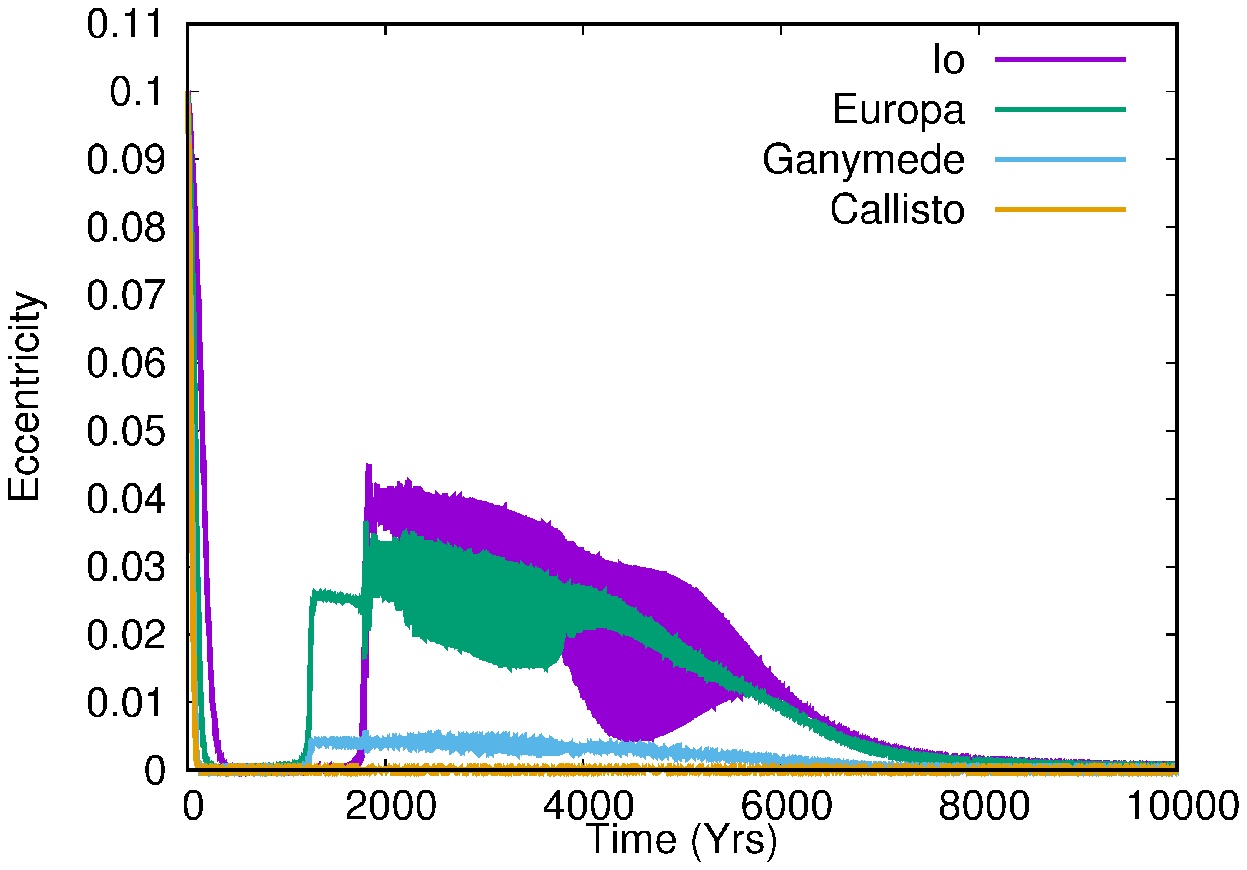}
  \includegraphics[scale=0.67]{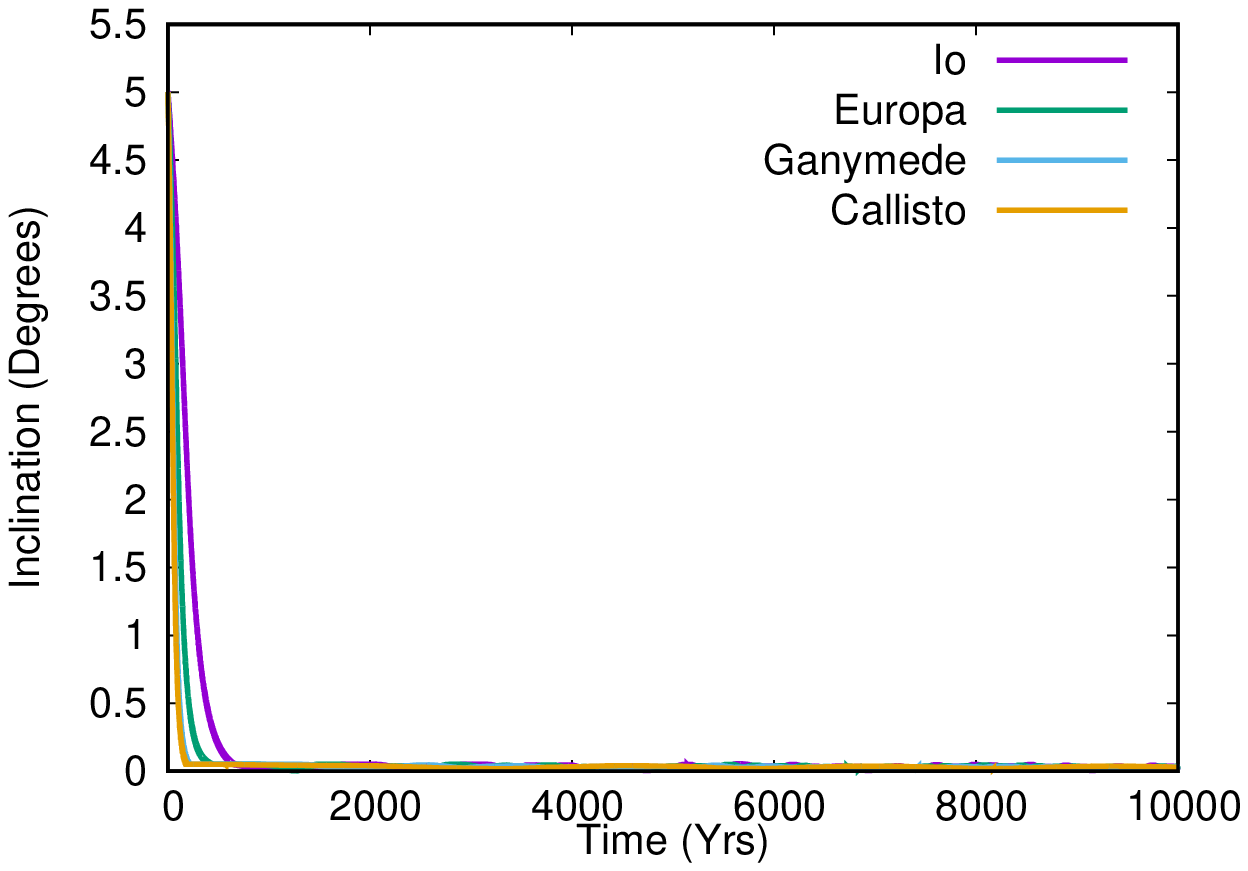}
  \caption[Semi-ecc.]{Evolution of eccentricity and inclination for the satellites for model \texttt{4-sat-2} over 10000~yrs. }
  \label{fig:inc-ecc-model-53} 
  \end{center} 
\end{figure}

To ensure that our results are independent of the assumption of initial circular and coplanar orbits of the satellites, we simulate the same system as before but considering the satellites in orbits with eccentricity of 0.1 and $5^{\circ}$ of inclination. The evolution of eccentricity and inclination are shown in Fig. \ref{fig:inc-ecc-model-53}.\\
The values of eccentricity and inclination are damped to almost zero in the first 1000 years by the damping scheme proposed in the section \ref{stwo}. After the damping, the satellites remained coplanar until the end of the simulation. The evolution of the eccentricity follows the results from model \texttt{4-sat-1}. No significant changes were found for the orbital migration of the satellites.\\
%%%%%%%%%%%%%%%%%%%%%%%%%%%%%%%%%%%%%%%%%%%%%%%%%%%%%%%%%%%%%%%%%%%%%%%%%%%%%%%%%%%%%%%%%%%%%%%%%%%%%%%%%%%%%%%%
\subsection{Formation of Four Satellites (Model \texttt{multi-sat-1})}
\begin{figure} 
  \begin{center}
  \includegraphics[scale=0.67]{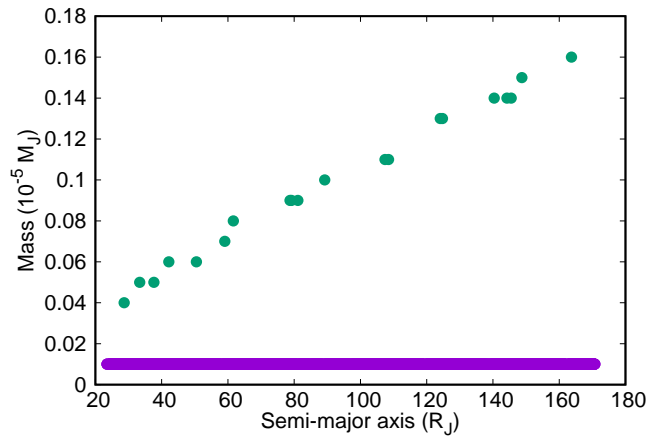}
  \caption[Semi-ecc.]{Initial mass distribution for model \texttt{multi-sat-1}.
  Green dots represent the satellite embryos and purple dots represent the sea of 2000 satellitesimals.}
  \label{fig:init-mass-model-nova-1-2} 
  \end{center} 
\end{figure}

Now that the migration and damping schemes were tested, we will move to simulations of forming satellites. First of all, the mass of the gaseous disc will be set to $2\times10^{-2}$ $M_J$, according to the MMSN model. For the solid disc, the mass distribution of individual satellite embryos scales as $r^{3/4}\Delta^{3/2}$ \citep{Kokubo-Ida-2002,Raymond-etal-2005,Izidoro-etal-2014} where $\Delta$ is the number of mutual Hill radii given by,
\begin{align}\label{eq:mutual-hill}
\Delta = \dfrac{a_i+a_j}{2}\left(\dfrac{M_i+M_j}{3M_p}\right)^{1/3},
\end{align}
with $a_i$, $a_j$, $M_i$ and $M_j$ being the semi-major axis and mass of two adjacent embryos. The mass of the satellitesimals is constant, $10^{-7}$ $M_J$. To reach the solid mass expected by the MMSN model we simulated 20 satellite embryos and 2000 satellitesimals randomly distributed from $21$ $R_J$ to $170$ $R_J$. In this model the eccentricity and inclination of all bodies are initially zero, the longitudes were randomly taken between $0^{\circ}$ and $360^{\circ}$ and the other orbital elements were set to zero. The initial mass distribution is shown in Fig. \ref{fig:init-mass-model-nova-1-2}, where the green dots are the satellite embryos and the purple dots are the satellitesimals. As one can see there is higher initial concentration of massive solids in the outer part of the disc ($r>30$ $R_J$), even though we did not specify the composition of the satellitesimals, the enhancement in the mass of the solids in the outer disc can be understood as due to ice condensation, however properly treatment for enhancement due to ice condensation will not be taken into account. Considering that the temperature has a radial dependence as $T\propto r^{-1}$ and the temperature for ice condensation is about 170 K, the ice line should be located around $4$ $R_J$, this topic will be addressed later. \par  
\citet{Miguel-Ida-2016} had shown that new generations of satellites are very unlikely to form, these authors adapted their model to add to the system a new embryo seed whenever an embryo collided with the central body, the mass of this embryo seed is 100 times lower than the original embryo and it is placed at the outer disc. They showed that the solid mass remaining in the disc is not enough to produce a new generation of satellites with mass comparable with the Jovian satellites. For this reason, we neglect the possibility of formation of new generations of satellites if the embryos collide with the planet.

\begin{figure} 
  \begin{center}
  \includegraphics[scale=0.67]{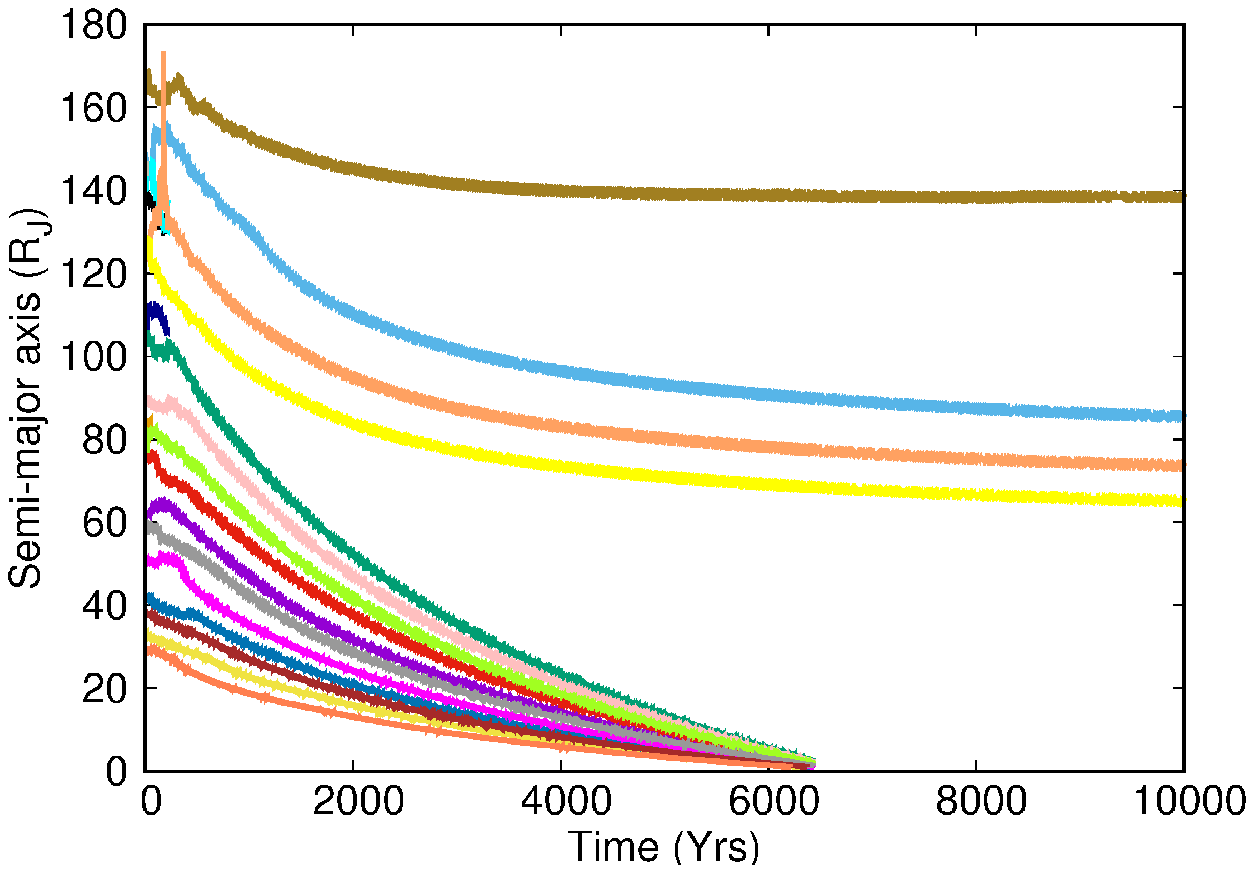}
  \includegraphics[scale=0.67]{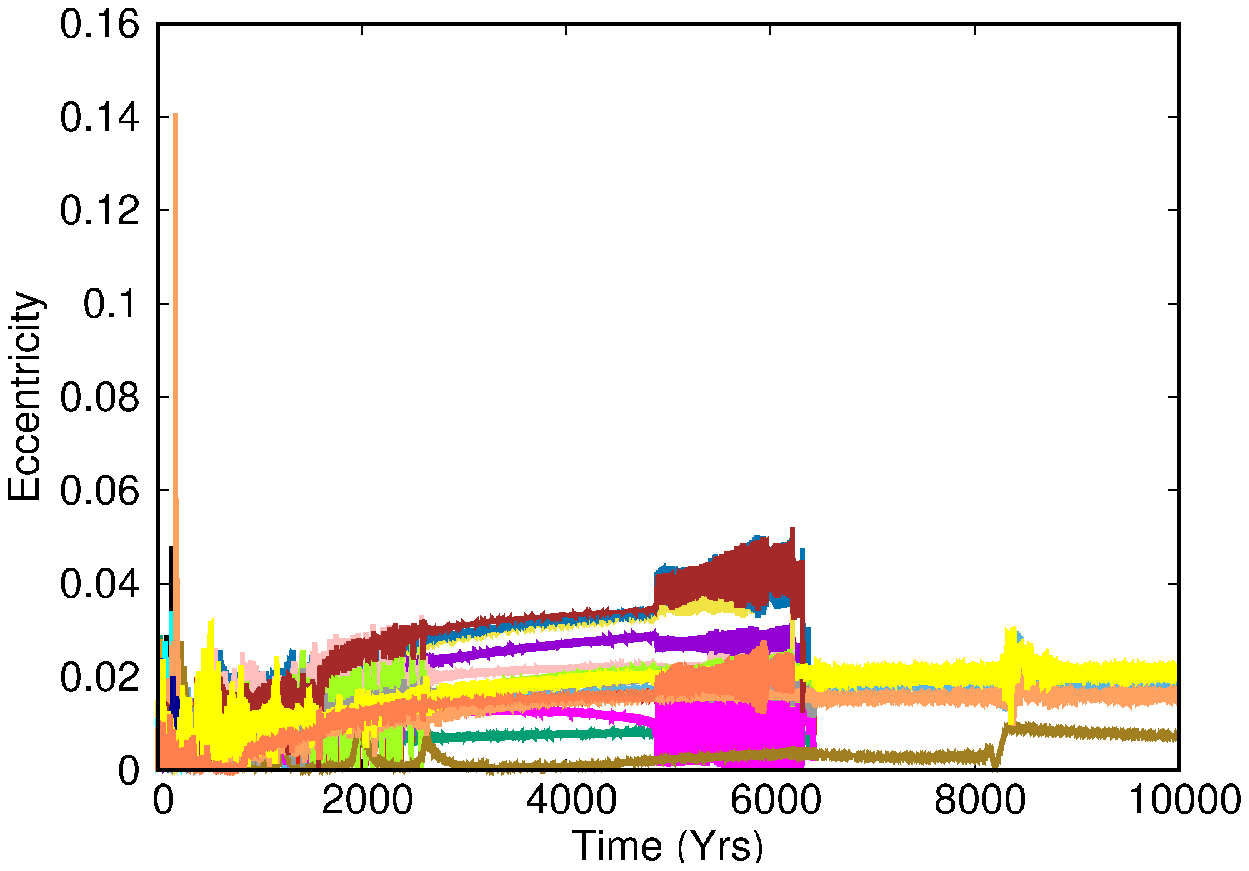}
  \caption[Semi-ecc.]{Evolution of the semi-major axis and eccentricity of the satellite embryos for model \texttt{multi-sat-1} over 10000~yrs.}
  \label{fig:semi-model-nova-1-2} 
  \end{center} 
\end{figure}  
The embryos migrate in the type I regime while the satellitesimals are dragged by aero-dynamical headwind described in the section \ref{stwo}. The evolution of the semi-major axis and eccentricity of the satellite embryos are shown in Fig. \ref{fig:semi-model-nova-1-2}.
The migration of the satellite embryos depends on the mass of the embryos and the density of the gaseous disc, in this way the embryos initially placed in the inner disc will migrate faster than the ones in the outer disc. Also, the inner embryos tend to accrete more material and become more massive than the outer satellite candidates. Those factors led to the scenario presented in Fig. \ref{fig:semi-model-nova-1-2} where the embryos located inside 120 $R_J$ migrated faster and collided with the central body while the outer embryos performed a low migration.\par
The inner cavity at 2.25 $R_J$ should prevent the embryos to fall immediately into the planet, however with many embryos reaching the inner cavity at, roughly, the same time, one embryo will push the other into the planet, causing a series of simultaneous collisions.\par
The surviving satellites came from the outer disc. Because of the initial density profile of the gaseous disc these bodies were formed in a region poor of gas, which explains the low migration rate presented by them. None the less, the outer bodies were not able to grow to massive satellites once the satellitesimals in the inner solid disc were accreted by the inner embryos or depleted by the gas drag and the interaction with the migrating bodies.

\begin{figure} 
  \begin{center}
  \includegraphics[scale=0.67]{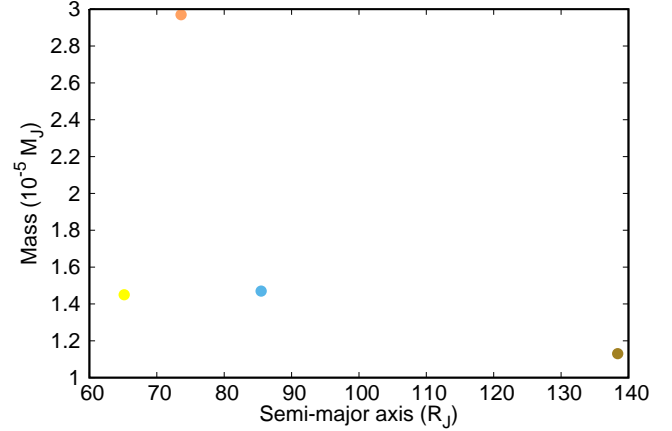}
  \caption[Semi-ecc.]{Mass versus semi-major axis of satellites formed due accretion of other embryos and satellitesimals for
  model \texttt{multi-sat-1}, after 10000~yrs.}
  \label{fig:mass-model-nova-1-2} 
  \end{center} 
\end{figure}

The satellite embryos grow through accretion of solids in their feeding zones, as the embryos migrate their respective feeding zones will move too, providing new material during the embryos evolution. In Fig. \ref{fig:mass-model-nova-1-2} we show the mass versus semi-major axis for the formed satellites. The inner embryos became massive faster and migrated toward the planet colliding with the central body. During the migration these satellites accreted most of the satellitesimals in the inner disc, also the aero-dynamical drag imposed on the satellitesimals contributed for the depletion of part of the solid disc. Because of these factors the outer embryos that survived are much less massive than the Galilean ones. \par
Another problem is the mass distribution of the formed satellites. In the Jovian system the two outer satellites are more massive than the inner ones, however we found it difficult to produce such a distribution in this first model. In our results the inner satellites are more massive following the trend that inner satellites grow faster and are more massive than the outer ones. 
%%%%%%%%%%%%%%%%%%%%%%%%%%%%%%%%%%%%%%%%%%%%%%%%%%%%%%%%%%%%%%%%%%%%%%%%%%%%%%%%%%%%%%%%%%%%%%%%%%%%%%%%%%%%%%%%
\subsection{Formation of More than Four Satellites  (Model \texttt{multi-sat-2})}
\begin{figure} 
  \begin{center}
  \includegraphics[scale=0.67]{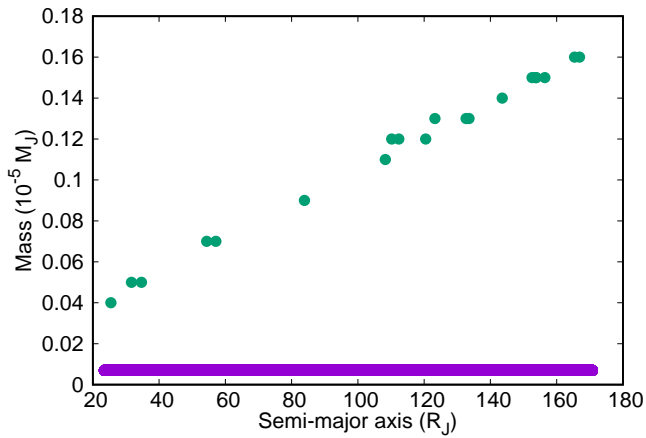}
  \caption[Semi-ecc.]{Initial mass distribution for model \texttt{multi-sat-2}. Green dots represent the satellite embryos and purple dots represent the satellitesimals.}
  \label{fig:init-mass-model-nova-2} 
  \end{center} 
\end{figure}

The MMSN model assumes the total mass of the solid disc, but how this mass is reached is one of the many uncertainties in the models of satellite formation. In the last model we opted for a configuration with 2000 satellitesimals and 20 embryos with masses chosen such as the gas to dust ratio will be 100. In this model, we will change the number and the mass of the satellitesimals while the number of embryos will remains the same. The embryos will be once again randomly distributed while the other parameters of the disc will be the same as in model \texttt{multi-sat-1}. We simulate 2500 satellitesimals with $7\times10^{-8}$ $M_J$ each (see fig. \ref{fig:init-mass-model-nova-2} for the initial distribution of the bodies). Our goal is to show how sensitive to the initial conditions this system can be.\par  
\begin{figure} 
  \begin{center}
  \includegraphics[scale=0.67]{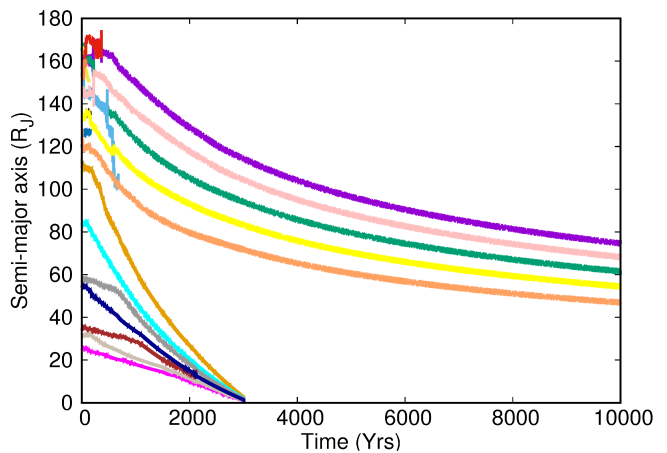}
  \caption[Semi-ecc.]{Evolution of the semi-major axis of the satellites embryos for model \texttt{multi-sat-2}, after 10000 years.}
  \label{fig:semi-model-nova-2-2} 
  \end{center} 
\end{figure}

At the end of the simulation five satellites were formed in orbits close to each other (Fig. \ref{fig:semi-model-nova-2-2}). A final configuration with more than four satellites is not so uncommon and was previously reported by \citet{Sasaki-etal-2010}, for example. As we know from our previous model, the number of formed satellites is strongly influenced by the initial distribution of the embryos, as only embryos initially located beyond 120 $R_J$ became satellites. In this model the random distribution of the embryos placed more bodies beyond 120 $R_J$ than in the model \texttt{multi-sat-1}, which might explain the formation of five satellites instead of four. We also performed simulations with the same distribution of embryos as in model \texttt{multi-sat-1}, changing the mass and number of satellitesimals (not showed here) and we found only four surviving satellites with configuration more compact than in the model \texttt{multi-sat-1}.\par
\begin{figure} 
  \begin{center}
  \includegraphics[scale=0.67]{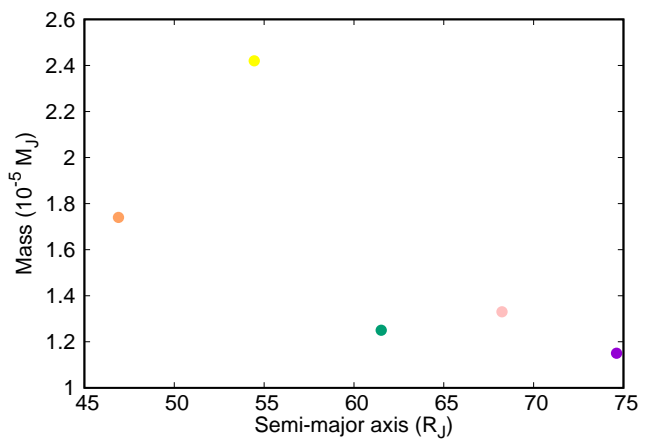}
  \caption[Semi-ecc.]{Mass versus semi-major axis of satellites formed due accretion of other embryos and satellitesimals for model \texttt{multi-sat-2}, after 10000 years.}
  \label{fig:mass-model-nova-2-2} 
  \end{center} 
\end{figure}

In Fig. \ref{fig:semi-model-nova-2-2} the orbital migration of the satellites embryos is shown. The general evolution is somehow similar to the one found in the model \texttt{multi-sat-1}. The embryos located at the inner regions of the disc grow faster through accretion of satellitesimals and they rapidly migrated toward the planet, after reaching the inner edge of the disc they collide with the central body. On the other hand, the migration of bodies from the outer disc is slower, because this is a region of low density and, at this point, the bodies are generally less massive than the ones from the inner disc. Comparing with the system formed in the model \texttt{multi-sat-1} the system in this model is more compact and is located inside 80 $R_J$, this configuration is a product of the initial distribution of the embryos and the interaction between the embryos in the outer disc. The initial number of satellitesimals is important only in the first phases of embryos growth, especially for the inner ones, these embryos grew faster than in the previous model and fell into the planet. For the outer embryos the effects of having more satellitesimals in the systems was not relevant.
The final mass distribution of the satellites is similar to the distribution presented in the model \texttt{multi-sat-1}, with the second innermost satellite being the most massive one and the outermost satellites being less massive (Fig. \ref{fig:mass-model-nova-2-2}). The increase in the number of satellitesimals was not enough to change this scenario, once the location of these new bodies was randomly selected. This tendency rises questions about how the Galileans reach their final masses, especially Ganymede and Callisto.         \par  
%%%%%%%%%%%%%%%%%%%%%%%%%%%%%%%%%%%%%%%%%%%%%%%%%%%%%%%%%%%%%%%%%%%%%%%%%%%%%%%%%%%%%%%%%%%%%%%%%%%%%%%%%%%%%%%%
\subsection{Formation of Several Satellites in Resonance  (Model \texttt{multi-sat-3})}
So far we have shown results when systems with four or five satellites  were formed, however the number of final satellites depends of the initial distribution of the embryos. We found that distributions with more embryos in the outer disc tend to produce systems with more satellites. The bodies formed in the previous models were not as massive as the Galileans and were located far from the regions of the regular satellites of Jupiter. In this model we will change the size of the disc and the initial mass of the satellite embryos. \par
\begin{figure} 
  \begin{center}
  \includegraphics[scale=0.67]{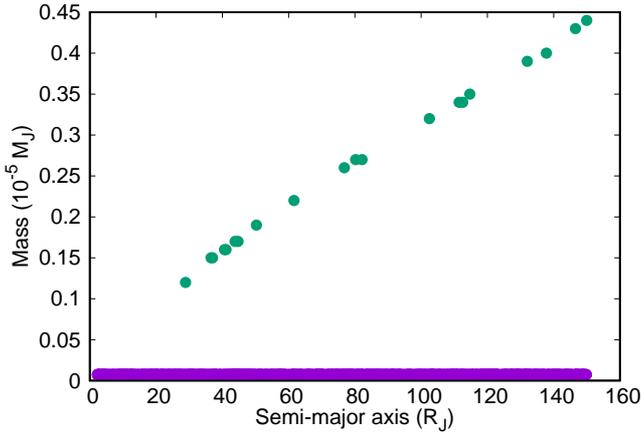}
  \caption[Semi-ecc.]{Initial mass distribution for model \texttt{multi-sat-3}. Green dots represent the satellite embryos and purple dots represent the satellitesimals.}
  \label{fig:init-mass-model-nova-3} 
  \end{center} 
\end{figure}

We will simulate 2000 satellitesimals with mass $7\times10^{-8}$ $M_J$, randomly spread from $2.25$ $R_J$ to $150$ $R_J$ and 20 satellite embryos with approximately three times the initial mass used in the previous two models, these bodies are distributed from $21$ $R_J$ to $150$ $R_J$. This new set-up is an attempt to solve the problem of formation of satellites with low mass at large distances. If the outer satellites become massive they could migrate to the inner regions of the disc. The initial solid mass distribution is presented in Fig. \ref{fig:init-mass-model-nova-3}, where the green points are the satellite embryos and the purple points are the satellitesimals. \par
\begin{figure} 
  \begin{center}
  \includegraphics[scale=0.67]{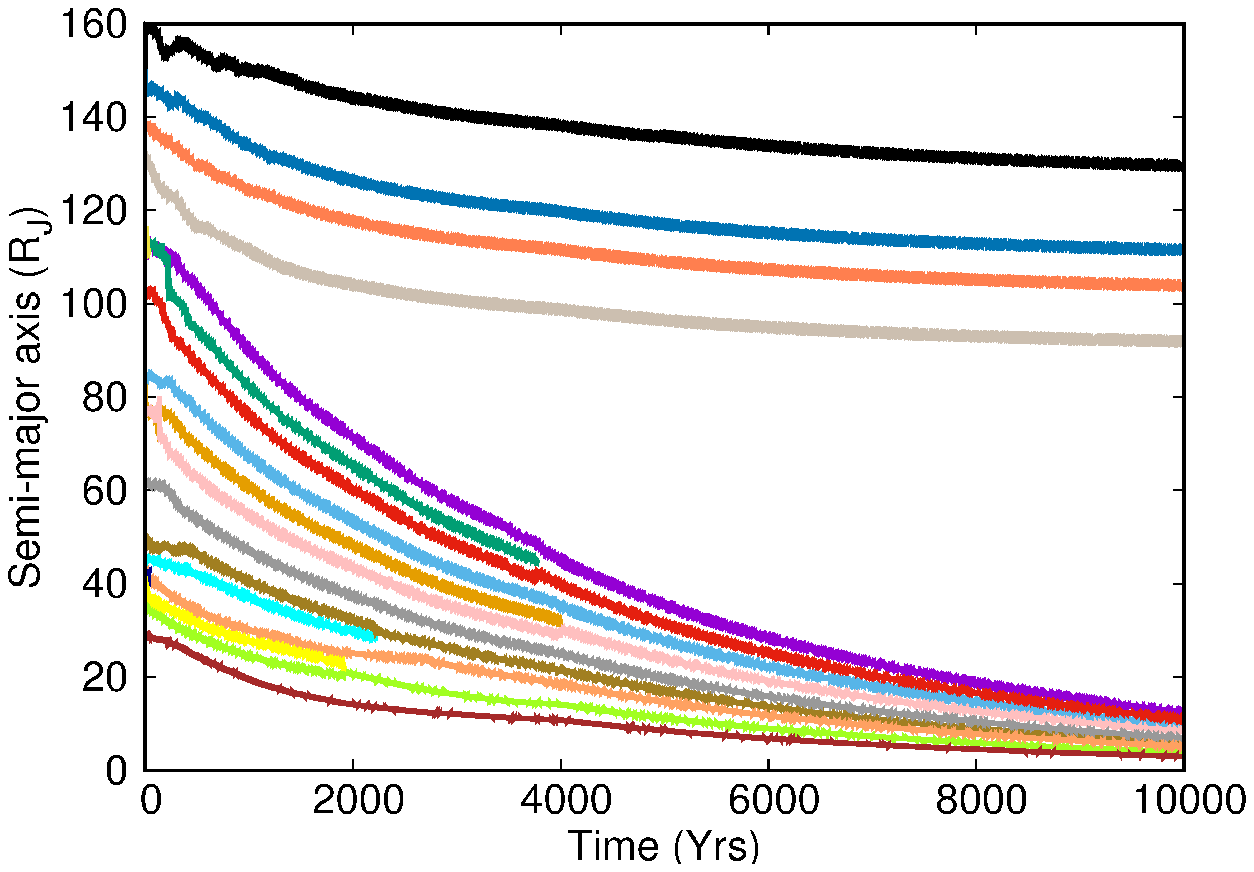}
  \includegraphics[scale=0.67]{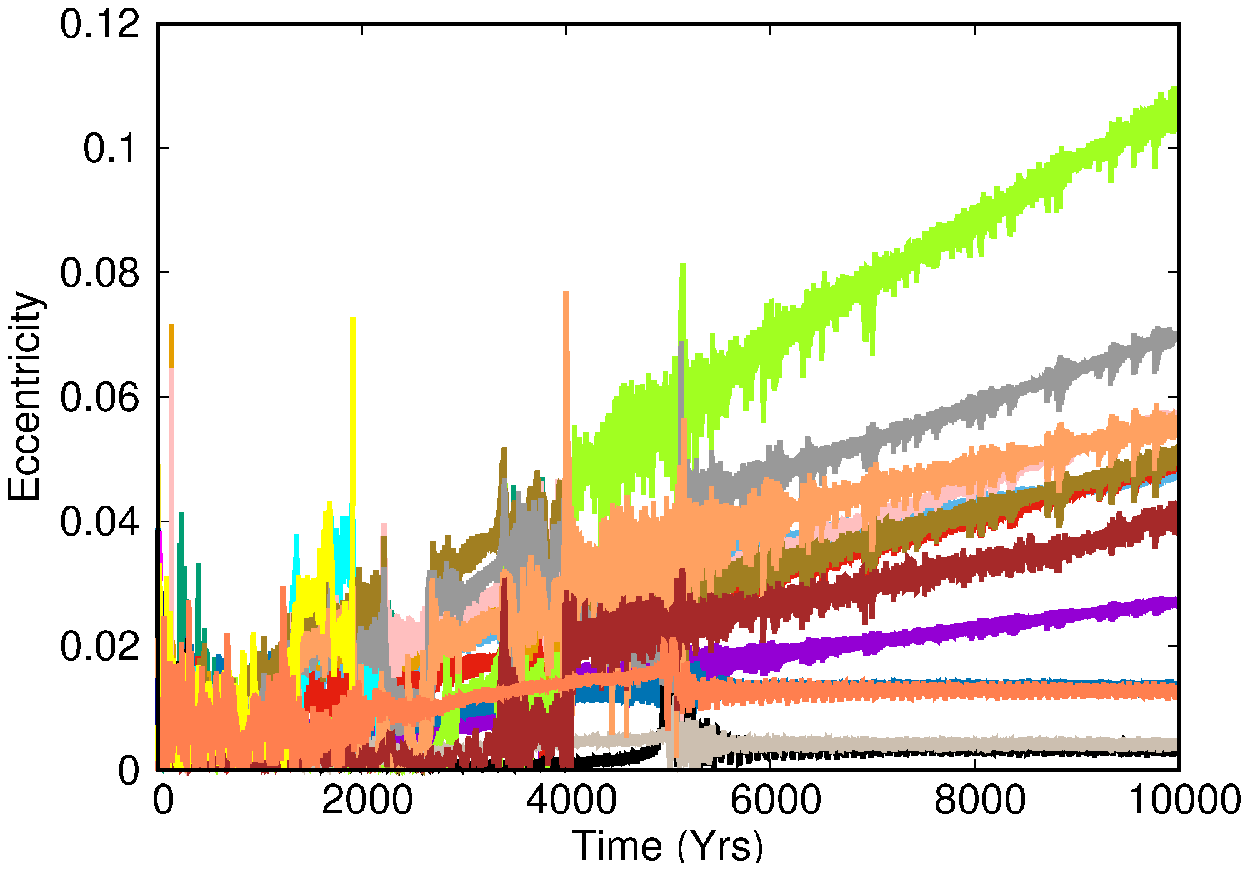}
  \caption[Semi-ecc.]{Evolution of the semi-major axis and eccentricity for the satellites for model \texttt{multi-sat-3} after 10000~yrs. }
  \label{fig:semi-ecc-model-nova-3} 
  \end{center} 
\end{figure}

The orbital evolution of the satellite embryos is shown in Fig. \ref{fig:semi-ecc-model-nova-3}. The migration of the embryos follows the trend of the previous model, the embryos inside 120 $R_J$ migrate faster and reach the inner edge of the disc, while the embryos located beyond 120 $R_J$ performed a slow migration. Different from the previous simulations the satellites did not collide with the central body, they piled up close to the inner cavity in a resonant configuration. In fact, when the inner cavity is placed at the co-rotation radius
of the planet the tendency is that a satellite halts its migration and does not fall into the planet. 
However the approach of multiple satellites almost at the same time might push the inner satellites toward the planet causing the lost of these bodies, as in the previous models. In this simulation, the migration speed of the satellites was lower compared to the models \texttt{multi-sat-1} and \texttt{multi-sat-2}, because most of the satellites inside $120$ $R_J$ entered in a resonant configuration instead of falling into the planet. After the resonant scenario be achieved the migration continued until the inner bodies reach the cavity.

Recent studies demonstrate that dissipation of the gaseous disc generates instabilities between close bodies in resonance, leading to the break-up of possible resonant chains \citep{Izidoro-etal-2017}. In our model the gaseous disc dissipates exponentially (equation \ref{eq:dens3}). Thus, the instabilities are expected and the resonant chains between the close satellites might not hold. In our results, the eccentricity of the inner satellites increases because of the compact configuration of these bodies (bottom panel of Fig. \ref{fig:semi-ecc-model-nova-3}), the instabilities generated by the dissipation of the gaseous disc also affects the satellites, however the resonant configuration is not broken until the end of our simulation. In this compact configuration the break-up of the resonant chains will produce several collisions between the satellites. A similar behaviour was observed for Super-Earth systems, where multiple planets piled up in resonant chains near the star, however, in this case, the instabilities due to the gas dissipation breaks these chains and the planets start to collide with each other or fall into the star \citep{Izidoro-etal-2017}.\par
The system of four satellites formed in the outer disc also presents a resonant configuration, however it is not expected that this configuration will be broken by instabilities due to gas dissipation. The gas density in the outer disc is already low and the dissipation of this disc will produce only weak instabilities which should not be strong enough to break the resonant chains between the satellites. This physical argumentation is supported by simulations involving Super-Earth systems in the literature.\par 
\begin{figure} 
  \begin{center}
  \includegraphics[scale=0.67]{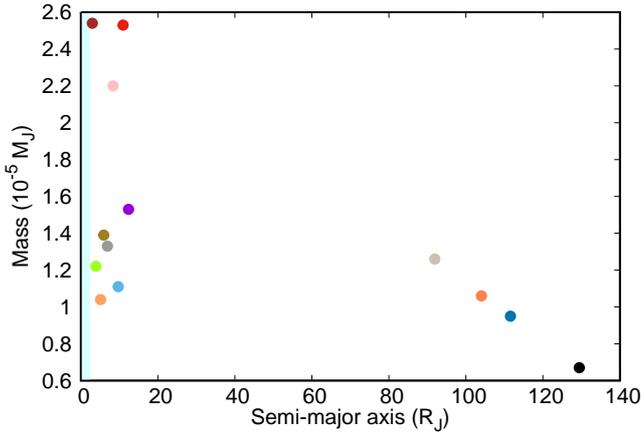}
  \caption[Semi-ecc.]{Mass versus semi-major axis of satellites formed due accretion of other embryos and satellitesimals for
  model \texttt{multi-sat-3}, after 10000~yrs. The cyan region represents the inner cavity of the disc.}
  \label{fig:semi-mass-model-nova-3} 
  \end{center} 
\end{figure}

In Fig. \ref{fig:semi-mass-model-nova-3} we show the final radial mass distribution of the satellites. It is possible to see the inner satellites piled up close to the inner cavity. As in the previous models most of the inner satellites are more massive than the outer ones.\par
The formation of numerous satellites shows the dependency on the initial conditions, in particular the radial distribution and the masses of the embryos. However, the biggest question that remains concerns the final masses of the satellites, so far all the formed systems have masses lower than the mass of the Galileans. Possible solutions include considering a solid and/or gas disc more massive than the predicted by the MMSN or change the density profile avoiding the fast migration of the inner embryos and decreasing the depletion of the satellitesimal disc.\par 

The final configuration of the system and the resonant chains formed during the evolution of the satellites resembles the evolution of Super-Earth planets around solar mass stars \citep{Izidoro-etal-2017}. From our findings it is possible to argue that satellite systems such as the Galileans might have a formation pathway similar to a planetary system with Super-Earths around a solar mass star on a different scale.
%
%%%%%%%%%%%%%%%%%%%%%%%%%%%%%%%%%%%%%%%%%%%%%%%%%%%%%%%%%%%%%%%%%%%%%%%%%%%%%%%%%%%%%%%%%%%%%%%%%%%%%%%%%%%%%%%%
\subsection{Growth and Migration of Satellites in a Uniform Gas Disc  (Models \texttt{multi-sat-uni-1})}
There are many uncertainties concerning the density profile of the gaseous disc, for simplicity we adopted a power law profile with index $\beta = 3/2$. This profile was chosen in order to have more material in the inner disc as it might be expected due to the inflow of material from the proto-planetary disc to the planet. However, the profile implies a fast migration of the inner embryos and slow migration of the outer ones, also the satellitesimals in the inner disc are depleted faster due to the high density in this region. For comparison we present in this section
a model in which we use an initial uniform gaseous disc with $\beta=0$. To ensure a fair comparison, the initial conditions of the embryos and the solids configuration will be the same as in model \texttt{multi-sat-1}.\par
With an initial uniform distribution of the gas it is expected that the migration rates of the satellite embryos to become more similar, decreasing the possibility of formation of satellite in the outer parts of the disc. Also, in the case of formation of resonant chains, the dissipation of the gaseous nebula might have a similar effect in the inner and outer disc.\par
\begin{figure} 
  \begin{center}
  \includegraphics[scale=0.67]{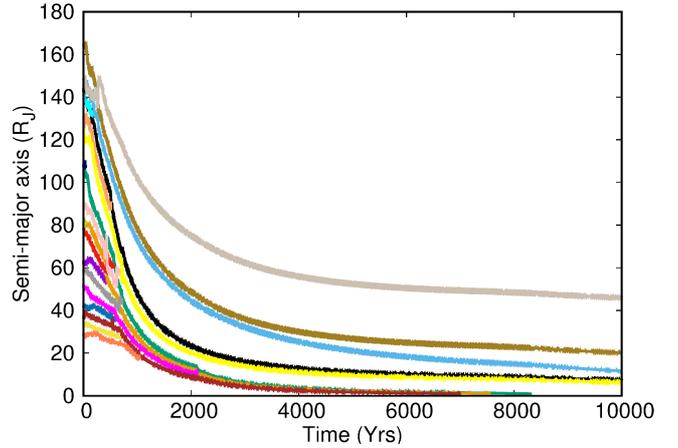}
  \caption[Semi-ecc.]{Evolution of the semi-major axis of the satellites embryos for the model \texttt{multi-sat-uni-1} after 10000 years.}
  \label{fig:semi-model-nova-1-3} 
  \end{center} 
\end{figure}

The orbital migration of the satellite embryos for our model is shown in Fig. \ref{fig:semi-model-nova-1-3}. Different from the models where the gas distribution has a radial dependence, now the embryos from the outer disc migrate faster changing the general migration of the inner embryos. We can notice that due to the migration of the outer embryos several collision took place well before 1000 years. At the end of the simulations, we have five surviving satellites, with the outermost one being inside $50$ $R_J$ while the other four are distribute from $6$$R_J$ to $27$ $R_J$. These final locations for the satellites reproduce the location of the Galilean satellites better than our previous models.\par
\begin{figure} 
  \begin{center}
  \includegraphics[scale=0.67]{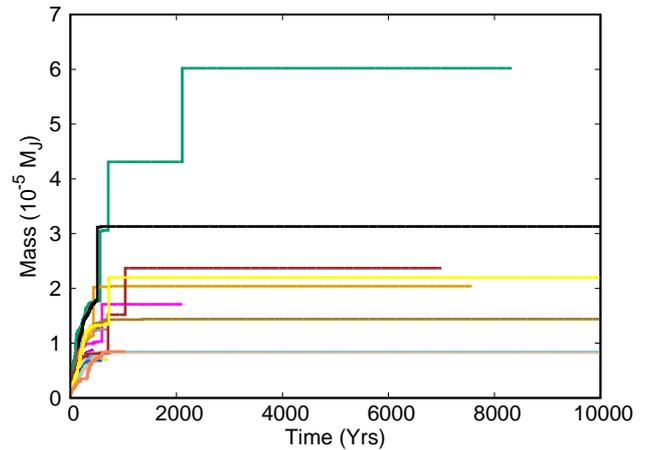}
  \caption[Semi-ecc.]{Evolution of the mass of the satellitesimals for the model \texttt{multi-sat-uni-1} after 10000 years.}
  \label{fig:mass-model-nova-1-3} 
  \end{center} 
\end{figure}

The final mass of the formed satellites remains an issue. In Fig. \ref{fig:mass-model-nova-1-3} we show the mass evolution of the embryos. Most of the satellitesimals are accreted in the first 500 years, the satellitesimals that are not accreted in this stage are scattered by the migrating embryos. The collision between embryos also contribute for the growth of the final satellites. After approximately 2000 years there is no much change in the masses of the remaining satellites for our model. The satellites in this models reached masses from $0.83\times 10^{-5}$ $M_J$ to $3.1\times 10^{-5}$ $M_J$. However, we can see in Fig. \ref{fig:mass-model-nova-1-3} that massive satellites, around $6\times 10^{-5}$ $M_J$, were formed during the evolution of the system, unfortunately this satellite was lost due to collision with the central planet. \par
Simulations with $\beta=1/2$ and $\beta=1.0$ were also performed. We decided not shown these results because in both cases the surviving satellites are located in the region between the satellites formed with $\beta=0$ and $\beta=3/2$ and are not as massive as the satellites formed with uniform distribution. These behaviours indicate that, under our assumptions, flatter gas discs are more favourable for the formation of satellites in the inner disc.
%%%%%%%%%%%%%%%%%%%%%%%%%%%%%%%%%%%%%%%%%%%%%%%%%%%%%%%%%%%%%%%%%%%%%%%%%%%%%%%%%%%%%%%%%%%%%%%%%%%%%%%%%%%%%%%%
\subsection{Models with Higher Aspect Ratio (Models \texttt{multi-sat-uni-1-h015}, \texttt{multi-sat-uni-2-h015} and \texttt{multi-sat-uni-3-h015})}
In the last section we approached the problem of fast migration of the innermost satellites changing the initial distribution of the gas density, the results shown an improvement in the migration but not in the masses of the final satellites. In this section we will address the questions about the temperature in the disc. \par
In our models we are not interested in develop a very sophisticated temperature profile, however from equation \ref{eq:temp1} we estimated that the ice line (when the temperature is $\le 170$ $K$) is around $4$ $R_J$, which is very close to the planet. The initial temperature of the disc is given by, 
\begin{align}\label{eq:t01}
T_0 = \dfrac{h^2GM_p\mu}{\mathcal{R}},
\end{align}
where $G$ is the gravitational constant, $\mu$ is the average molecular weight and $\mathcal{R}$ is the gas constant. Thus, the initial temperature can be changed modifying the value of the aspect ratio $h$, consequently is possible to initially place the ice line in a more reliable location. Following the work by \citet{Lunine-Stevenson-1982} and \citet{Miguel-Ida-2016}, for example, the ice line is predict to be located around $30$ $R_J$, to match this value we set the aspect ratio as $h=0.15$ (an intermediate value could be chosen for the aspect ratio, lets say 0.1, however it will not move the ice line to $30$ $R_J$). As a consequence for increase the value of $h$, the migration rate of the satellites should decrease because the disc becomes thicker, in this way, and based in the results from last section, we opt to use an uniform gas density profile. Our simulations will have the same initial conditions as the models \texttt{multi-sat-1}, \texttt{multi-sat-2} and \texttt{multi-sat-3}, only changing the gas density profile and the aspect ratio, as specified before. To avoid complications, we will name our new models as follows, the model \texttt{multi-sat-uni-i-h015} will have the same initial configuration as model \texttt{multi-sat-i} with an initial uniform distribution for the gaseous disc and a $h=0.15$, where $i=1,2,3$.\par
\begin{figure} 
  \begin{center}
  \includegraphics[scale=0.67]{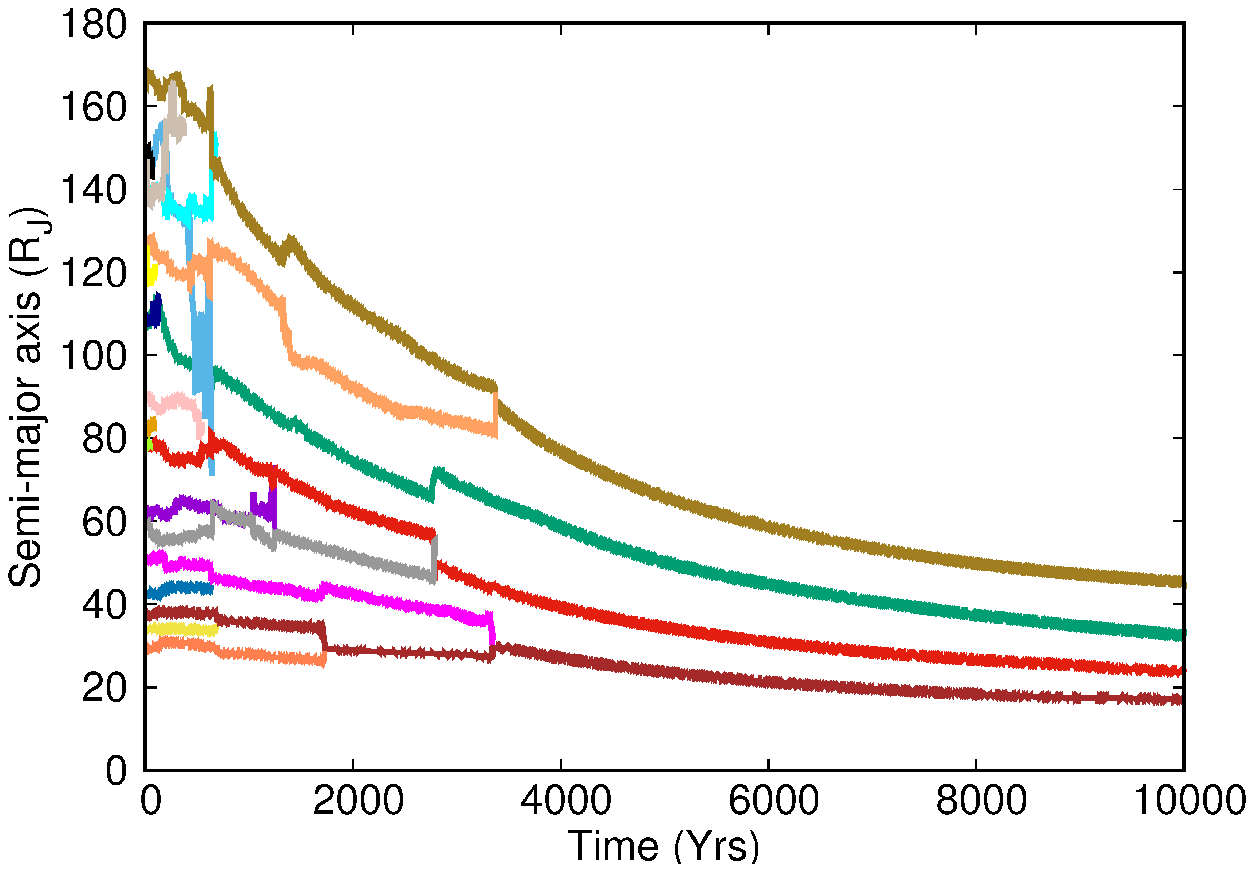}
  \includegraphics[scale=0.67]{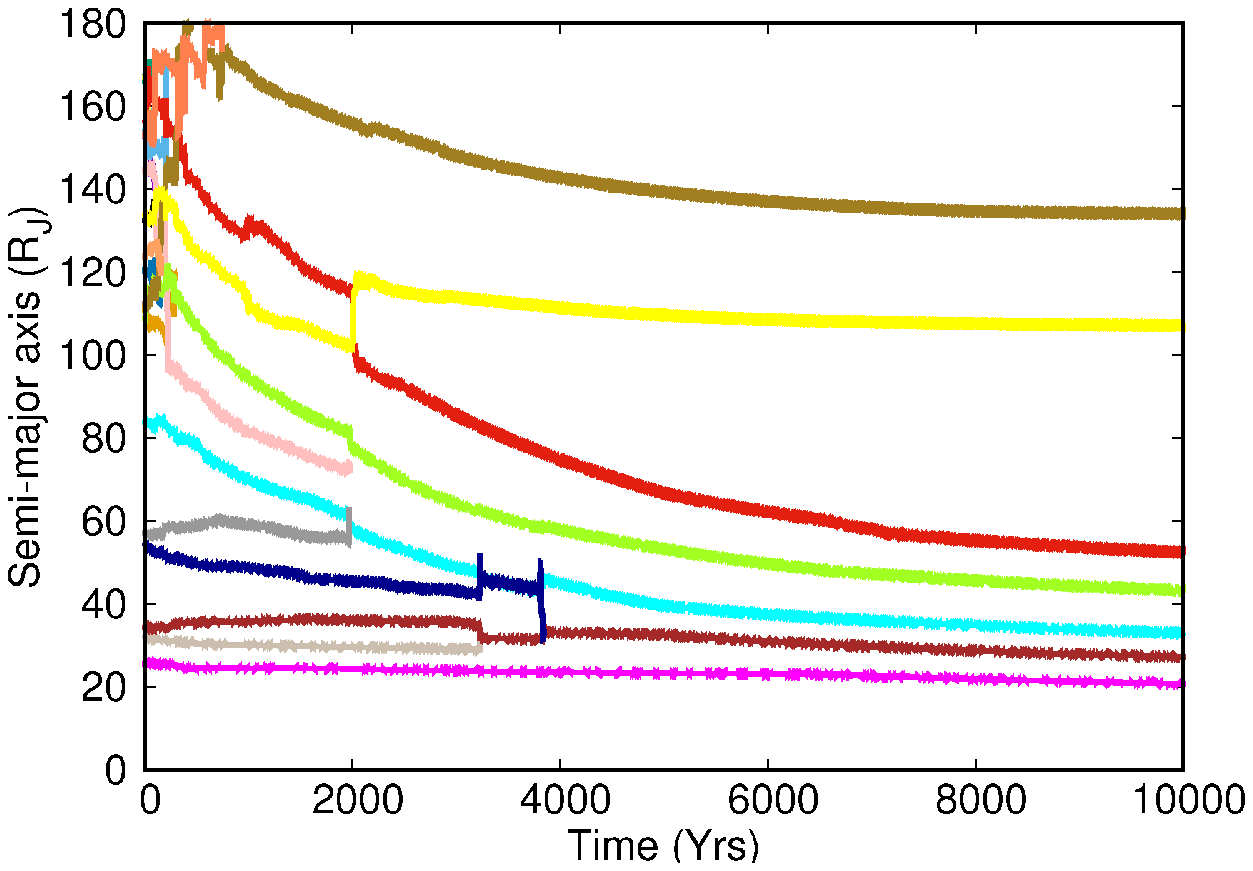}
  \includegraphics[scale=0.67]{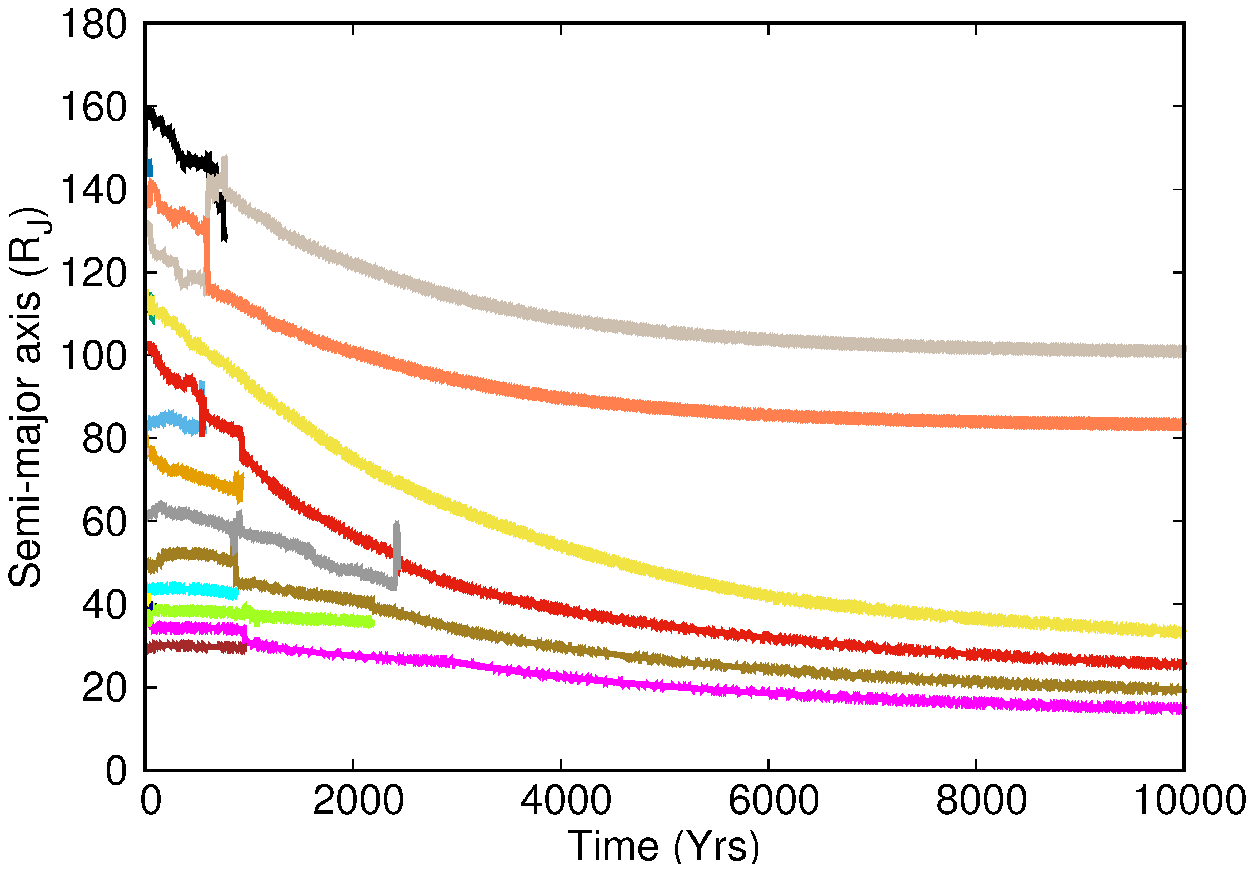}  
  \caption[Semi-ecc.]{Evolution of the semi-major axis of the satellites embryos for the models \texttt{multi-sat-uni-1-h015}, \texttt{multi-sat-uni-2-h015} and \texttt{multi-sat-uni-3-h015} after 10000 years, respectively from top to the bottom.}
  \label{fig:semi-model-nova-1-3-h015} 
  \end{center} 
\end{figure}

Fig. \ref{fig:semi-model-nova-1-3-h015} shows the evolution of the semi-major axis of the embryos for the models \texttt{multi-sat-uni-1-h015}, \texttt{multi-sat-uni-2-h015} and \texttt{multi-sat-uni-3-h015}. As expected, the migration rate of the embryos decreased compared with the model \texttt{multi-sat-uni-1}, for all three models the inner satellites are distribute from around $20$ $R_J$ to $60$ $R_J$, while the outer ones were formed beyond $80$ $R_J$. Similarly to model \texttt{multi-sat-uni-1} the evolution of the embryos here was determined by multiple collisions between each other, specially in model \texttt{multi-sat-uni1-h015} where only four satellites were formed with no first order resonance holding them together. For the other two models (\texttt{multi-sat-uni-2-h015} and \texttt{multi-sat-uni-3-h015}), we have four and five satellites forming in the inner group and two satellites forming in the outer group, also a small number of first order resonance chains were found for the satellites located in the inner group in both models.\par 
\begin{figure} 
  \begin{center}
  \includegraphics[scale=0.67]{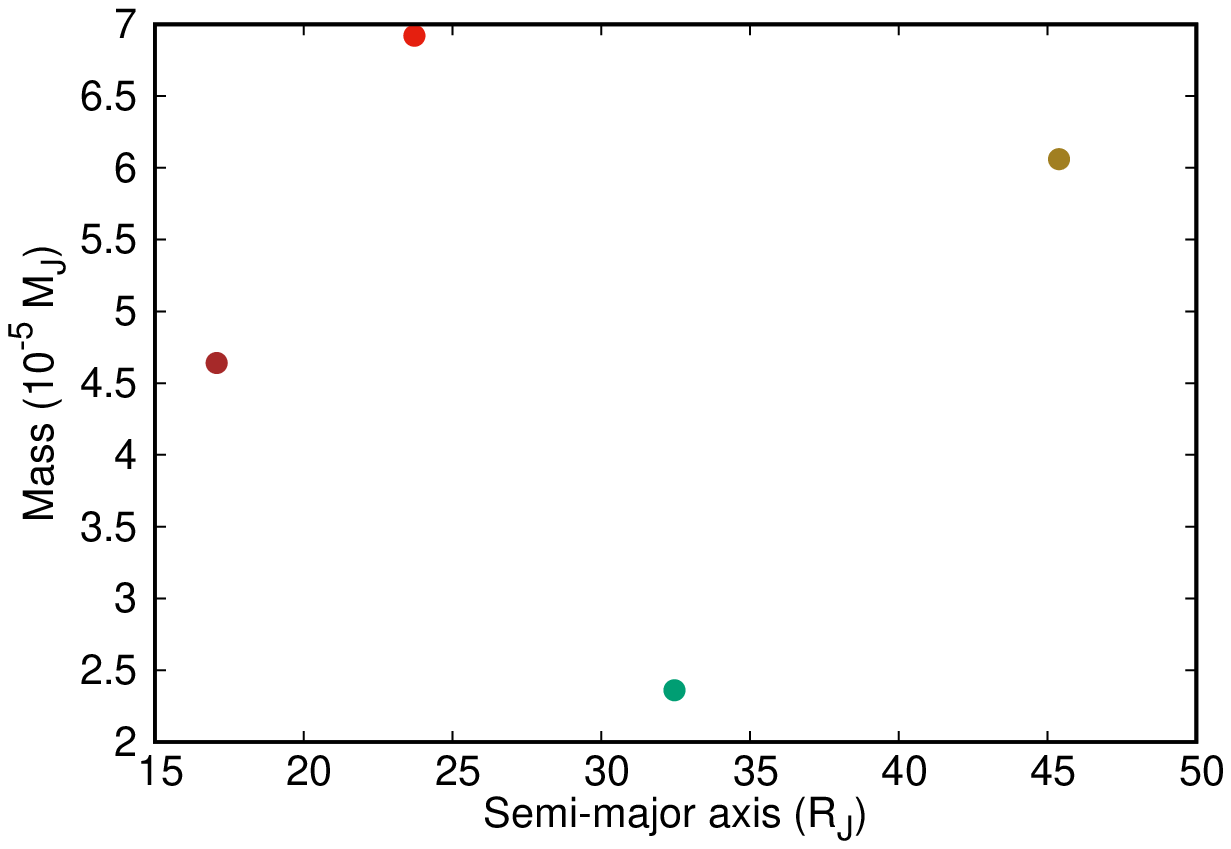}
  \includegraphics[scale=0.67]{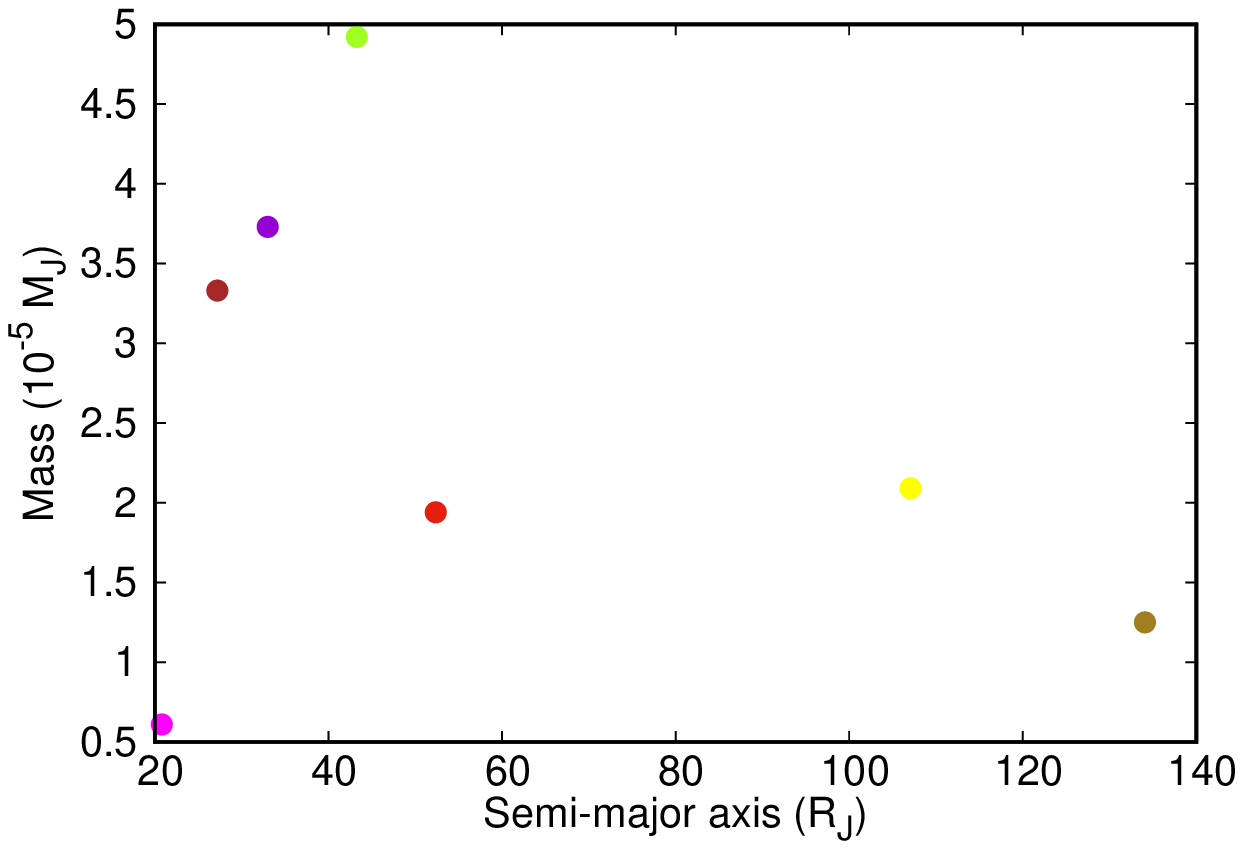}
  \includegraphics[scale=0.67]{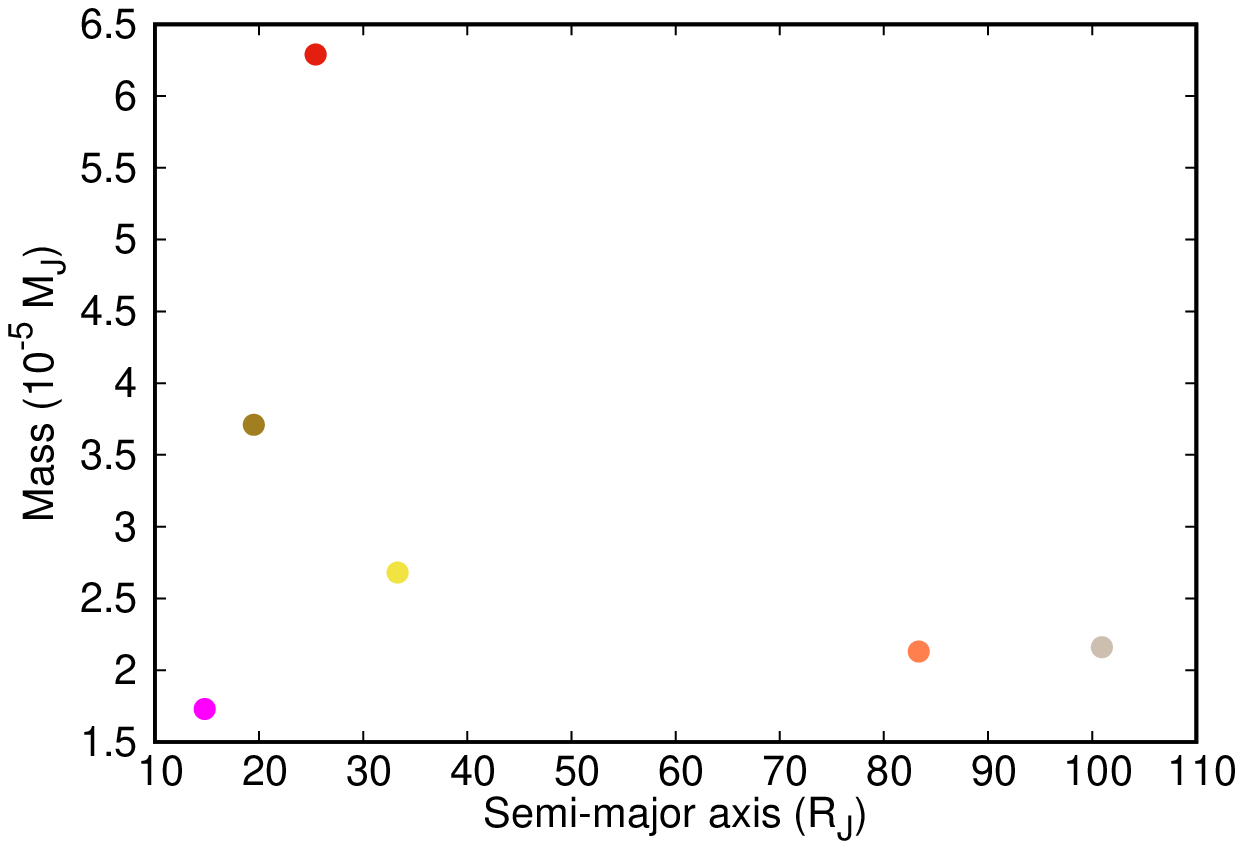}
  \caption[Semi-ecc.]{Mass versus semi-major axis of satellites formed due accretion of other embryos and satellitesimals for  models \texttt{multi-sat-uni-1}, \texttt{multi-sat-uni-2} and \texttt{multi-sat-uni-3} after 10000 years, respectively from top to bottom.}
  \label{fig:mass-model-nova-1-3-015} 
  \end{center} 
\end{figure}

The final radial distribution of mass for our models is shown in Fig. \ref{fig:mass-model-nova-1-3-015}. We highlight the enhancement in the masses of the formed satellites, in the models \texttt{multi-sat-uni-1-h015} and \texttt{multi-sat-uni-3-h015} the most massive satellite is almost as massive as Ganymede while in the model \texttt{multi-sat-uni-2-h015} the most massive satellite has a mass comparable to Io. These models are the only models presented so far that produced satellite with masses comparable with the Galileans.\par
The model \texttt{multi-sat-uni-1-015} was capable to reproduce the masses of the Galileans satellites very accurately, where Io was reproduced by the first satellite, Ganymede was reproduced by the second one, Europa was reproduced by the third one and Callisto was reproduced by the fourth satellite (ordering our satellites from the innermost to the outermost). The only issue are the switched position between the Ganymede analogous and the Europa analogous, and the fact that our group of satellite is not in the exactly location of the Galilean satellites. 
Even thought the position of the formed satellites was not well reproduced, the increase in the aspect ratio of the disc was crucial for the improvement in the mass of the satellites, here we could reproduce the masses of the Galileans and see that our approach to the problem can generate massive satellites once the position of the ice line is moved to around $30$ $R_J$. In this way, we will use this same aspect ratio and gas density distribution for the upcoming models.  

%%%%%%%%%%%%%%%%%%%%%%%%%%%%%%%%%%%%%%%%%%%%%%%%%%%%%%%%%%%%%%%%%%%%%%%%%%%%%%%%%%%%%%%%%%%%%%%%%%%%%%%%%%%%%%%%
\subsection{Model with an Enhanced Solids Composition (Model \texttt{multi-sat-uni-4-h015})}
\label{subsec:enhanced}
In our previous models we changed the number and mass of embryos and satellitesimals always keeping the gas to dust ratio to be 100. However, different ratios can be tested under the MMSN assumptions. Here we decide to test a model with a higher concentration of solids, in this way we will build a disc with gas to dust ratio of 10. \par
We will use the model \texttt{multi-sat-uni-1-h015} as a reference for the initial conditions. In order to enhance the concentration of solids, we preserve the original amount of solids and remove a certain amount of gas from the disc until the gas to dust ratio becomes 10. In this way we are still considering a minimum mass disc model for solids. Differently to \citet{Mosqueira-Estrada-2003b} and \citet{Estrada-etal-2009}, we choose to decrease the gas density in all parts of the disc equally, not giving preference for a specific region.\par
\begin{figure} 
  \begin{center}
  \includegraphics[scale=0.67]{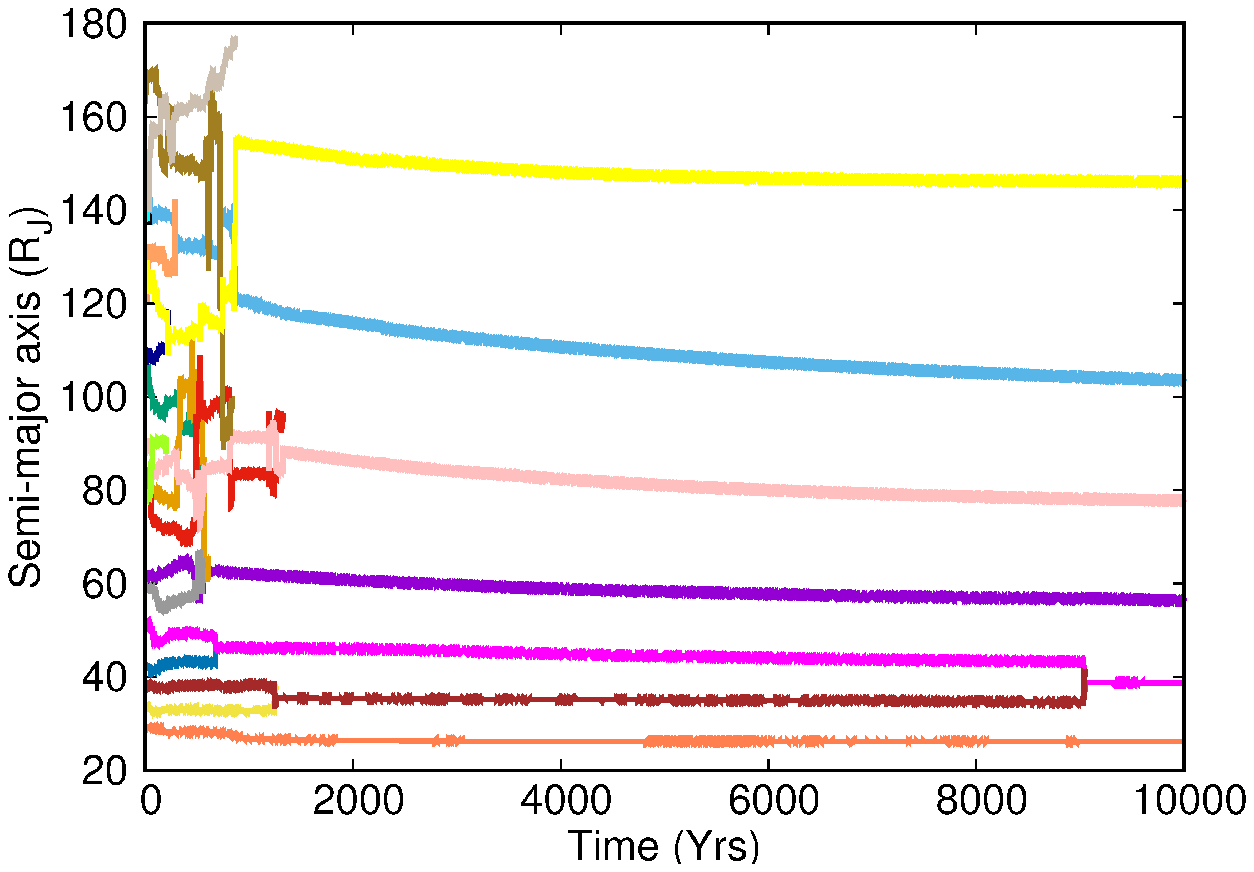}
  \includegraphics[scale=0.67]{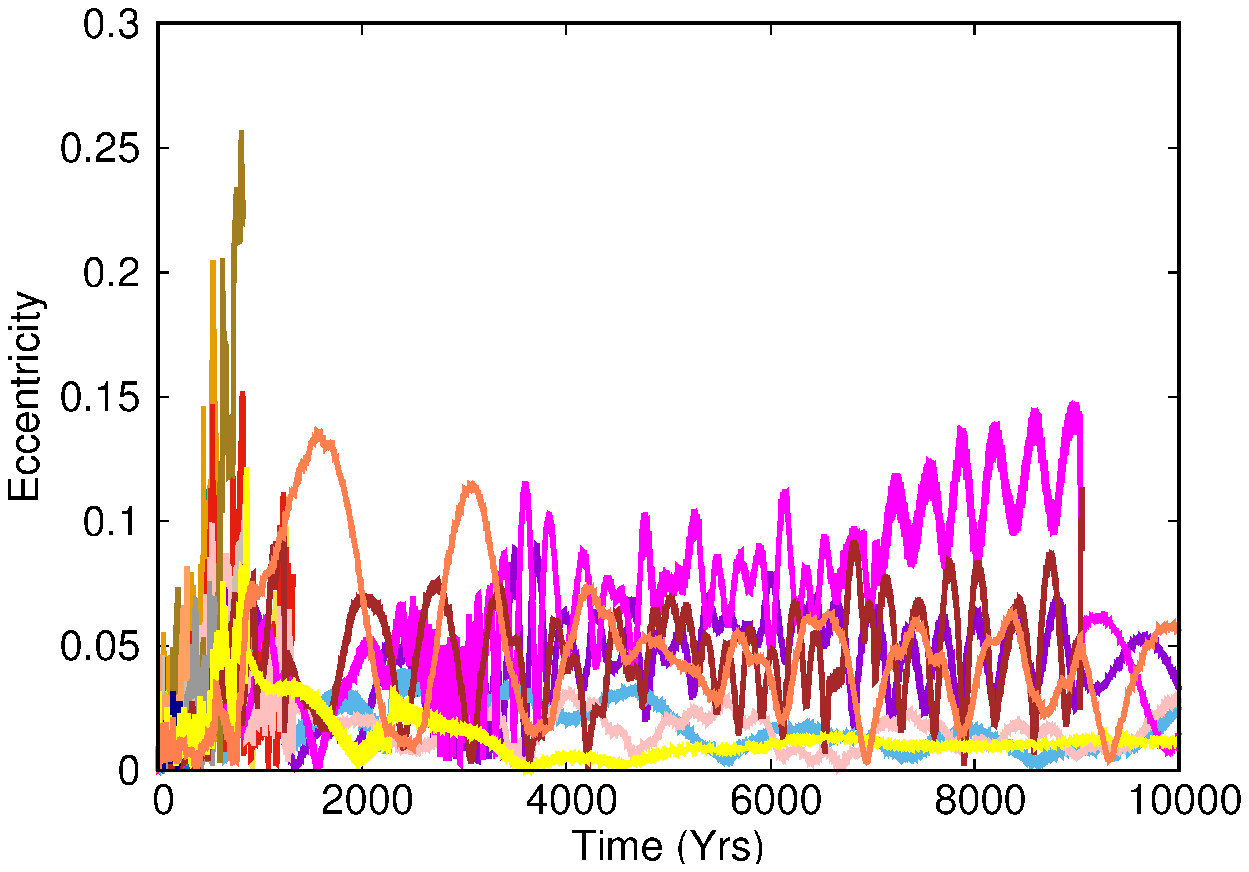}
  \caption[Semi-ecc.]{Evolution of the semi-major axis and eccentricity for the satellites for model \texttt{multi-sat-uni-4} after 10000~yrs. }
  \label{fig:semi-ecc-model-nova-1-3-gas-dust-10} 
  \end{center} 
\end{figure}

In Fig. \ref{fig:semi-ecc-model-nova-1-3-gas-dust-10} we show the evolution of semi-major axis and eccentricity for the model \texttt{multi-sat-uni-4-h015}. The reduction of the gas in the nebula slow the migration rate of the embryos, such that the number of satellites formed is higher compared with the model \texttt{multi-sat-uni-1-h015}, also the formed satellites are located farther away in the disc. Only one 3:2 resonance chain was found locking the fourth and fifth outermost satellite.\par 
\begin{figure} 
  \begin{center}
  \includegraphics[scale=0.67]{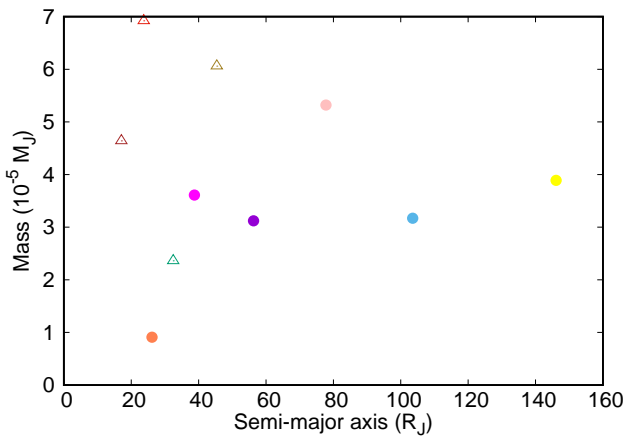}
  \caption[Semi-ecc.]{Comparison of the final mass distribution of satellites formed due accretion of other embryos and satellitesimals for the 
  models \texttt{multi-sat-uni-4-h015} (filled circles) and \texttt{multi-sat-uni-1-h015} (asterisk marks), after 10000~yrs. }
  \label{fig:semi-mass-model-nova-1-3-gas-dust-10} 
  \end{center} 
\end{figure}

Fig. \ref{fig:semi-mass-model-nova-1-3-gas-dust-10} shows the comparison between the final mass distribution of the satellites for the models \texttt{multi-sat-uni-4-h015} and \texttt{multi-sat-uni-1-h015}. We can see that the values of masses for the model \texttt{multi-sat-uni-4-h015} are slightly smaller than in the model \texttt{multi-sat-uni-1-h015}, probably because of the formation of more satellites. The masses of the satellites in this model are more compatible with the masses of model \texttt{multi-sat-uni-2-h015}.
%%%%%%%%%%%%%%%%%%%%%%%%%%%%%%%%%%%%%%%%%%%%%%%%%%%%%%%%%%%%%%%%%%%%%%%%%%%%%%%%%%%%%%%%%%%%%%%%%%%%%%%%%%%%%%%%
\subsection{Model with a Uniform Gas disc and an Steep Distribution for the Embryos (Model \texttt{multi-sat-uni-5-h015})}
\begin{figure} 
  \begin{center}
  \includegraphics[scale=0.67]{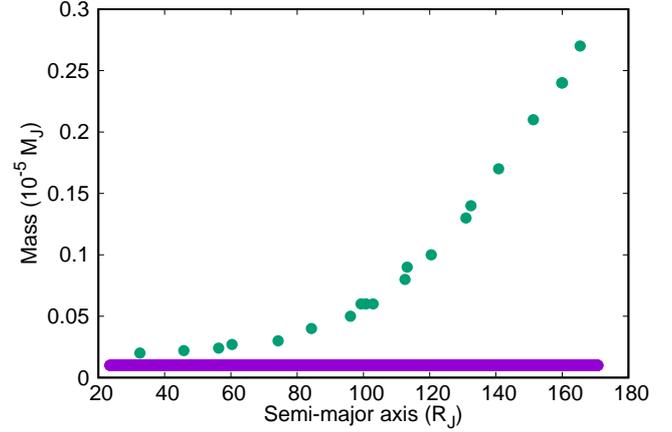}
  \caption[Semi-ecc.]{Initial mass distribution for model \texttt{multi-sat-uni-5}. Green dots represent the satellite embryos and purple dots represent the satellitesimals.}
  \label{fig:mass-init-model-uni-5} 
  \end{center} 
\end{figure}

So far we treated the embryos distribution such as it scales as $r^{3/4}\Delta^{3/2}$, however this distribution depends on the slope of the surface density distribution. In this way, for uniform distributions as discussed in the models \texttt{multi-sat-uni-1} and \texttt{multi-sat-uni-i-h015}, the distribution of the embryos should scale differently. To cover this, in this section we will run a simulation with the same set up as model \texttt{multi-sat-uni-1-h015} with the embryos distribution proportional to $r^{3}$ \citep{Kokubo-Ida-2002,Raymond-etal-2005,Izidoro-etal-2014}. The initial distribution of the embryos (and the satellitesimals) is shown in Fig. \ref{fig:mass-init-model-uni-5}, as the distribution is steeper, the difference in mass between the more massive and the less massive embryo is greater than found in the previous models. \par
\begin{figure} 
  \begin{center} 
  \includegraphics[scale=0.67]{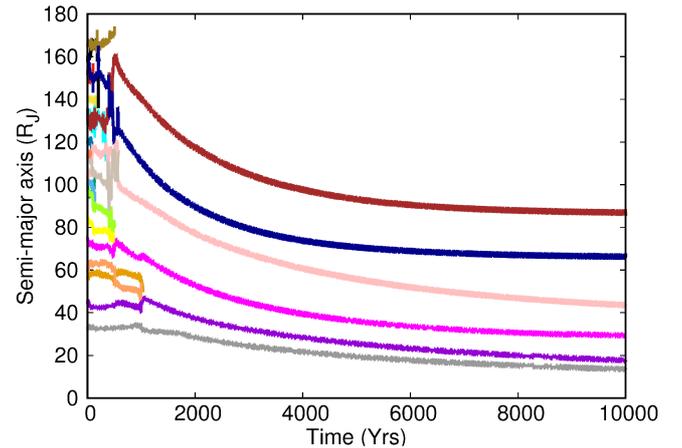}
  \caption[Semi-ecc.]{Evolution of the semi-major axis for model \texttt{multi-sat-uni-5-h015} after 10000~yrs. }
  \label{fig:semi-model-uni-5} 
  \end{center} 
\end{figure}

The migration of the embryos followed the pattern found for the models with uniform surface density, six satellites survived inside $13.5$ $R_J$ and $87$ $R_J$ (Fig. \ref{fig:semi-model-uni-5}).  Only two 3:2 resonant chains were found, which is the most common commensurability in our simulations. As we pointed before, the initial distribution of the embryos changes the final number of satellites, however their final location is an outcome of the gas density profile and the scale height.  \par
\begin{figure} 
  \begin{center}
  \includegraphics[scale=0.67]{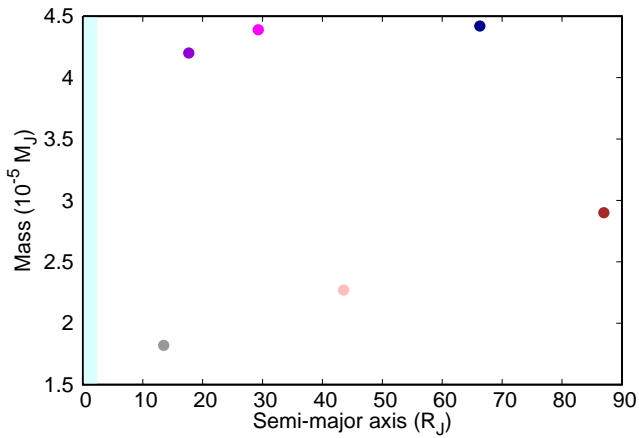}
  \caption[Semi-ecc.]{Mass versus semi-major axis of satellites formed due accretion of other embryos and satellitesimals for
  model \texttt{multi-sat-uni-5-h015}, after 10000~yrs. The cyan region represents the inner cavity of the disc. }
  \label{fig:mass-model-uni-5} 
  \end{center} 
\end{figure}

Fig. \ref{fig:mass-model-uni-5} shows the final distribution of mass of the surviving satellites. Following the trend of the models with uniform surface density, the satellites are more massive than the ones from simulation with $\beta=3/2$ and $h=0.05$. Comparing exclusively with the model \texttt{multi-sat-uni-1-h015}, the final masses of the satellites are smaller, even though some of these satellites started with a higher initial mass and the process of capture of satellitesimals and small embryos was easier, probably the formation of more than four satellites was the reason they do not became more massive. None the less, is important to notice that even the most massive satellite produced in this model is only comparable to Io, highlighting the difficulties to create satellites as massive as Ganymede and Callisto.

%%%%%%%%%%%%%%%%%%%%%%%%%%%%%%%%%%%%%%%%%%%%%%%%%%%%%%%%%%%%%%%%%%%%%%%%%%%%%%%%%%%%%%%%%%%%%%%%%%%%%%%%%%%%%%%%
\subsection{Close-in Super-Earths Systems vs Multiple Satellites Systems}
The comparison between the formation of the Galileans system with the formation of planets in our Solar System, is not something new, however the better match may be found among the extrasolar planets. The close-in Super-Earths are planets with masses of $\sim 10^{-5} - 10^{-4}$ $M_{\odot}$ orbiting close to the host star, with orbital periods shorter than 100 days \citep{Mayor-etal-2011, Howard-etal-2012, Petigura-etal-2013}. It is believed that the close-in Super-Earths were formed in a circum-stellar disc and only accrete solid materials analogous to the Galilean satellites, in addition the Super-Earths have the same mass ratio as the Jovian satellites \citep{Sasaki-etal-2010}.\par
Recently, the studies about the formation of close-in Super-Earths received a boost with the confirmation of seven Earth-size planets with very short orbital periods in the TRAPPIST-1 system \citep{Gillon-etal-2016, Gillon-etal-2017}, some of them in resonance \citep{Luger-etal-2017}. Studies, such as \citet{Ormel-etal-2017} and \citet{Izidoro-etal-2017}, were performed looking for a better understanding about formation and evolution of Super-Earths systems in general, using resonant chains as a constraint for the problem.\par 
The model proposed by \citet{Izidoro-etal-2017} showed the possibility of formation of multiple planets close to the star in resonance and how this resonance could be broken due to instabilities. Comparing their results with ours, in particular model \texttt{multi-sat-3}, we can see that the formation of resonant chains in systems with multiple Super-Earths is also found in systems with multiple satellites close to each other. \citet{Izidoro-etal-2017} pointed that the dissipation of the gaseous nebula induces instabilities in the planetary orbits which breaks the resonant chains leading to several collisions between the planets. This behaviour was not found for satellites, once the evolution of the system was finished with the satellites locked in resonances. The final resonant configurations of the satellites might be a consequence of the small mass of these bodies. Also, the high eccentricity presented by the Super-Earths in resonance was similar for the satellites. \par 
Despite the general similarities in the models and results, we stress that for Super-Earths systems the turbulence in the disc plays an important role, while for satellites the disc is usually considered to be laminar.\par 
In this section we are comparing only the dynamical evolution of satellites and Super-Earths, we do not claim that every feature of the formation process is the same. It is known that the composition of Super-Earths and satellites can be very different, which means that their formation discs were different, however the inward migration, the formation close to the central body and the resonant chains indicate that they experienced similar evolutions. In this way, the results for satellites in a circum-planetary disc can be scaled for Super-Earths in a proto-planetary disc.\par 
%%%%%%%%%%%%%%%%%%%%%%%%%%%%%%%%%%%%%%%%%%%%%%%%%%%%%%%%%%%%%%%%%%%%%%%%%%%%%%%%%%%%%%%%%%%%%%%%%%%%%%%%%%%%%%%%
Table \ref{tab:sumary} shows a quick summary of our result with the number of satellites formed, range of final semi-major axis, mass and eccentricity of the satellites and the number of satellites in first order resonances.

\begin{table*}
\caption[Models]{Sumary of our results. Where $a$, $M$ and $e$ are the range of values for the semi-major axis, mass and eccentricity of the satellites and in the last column the number of satellites in first order resonances (2:1, 3:2, 4:3, 5:4, 6:5 and 7:6).}
\begin{tabular}{c|c|c|c|c|c}\hline 
Models	    	 			 & Satellites   &   $a$				& $M$ 			& $e$  				& $1^{st}$ order MMR			\\
							 & 	  		  	&	$R_J$ 			& $10^{-5} M_J$	& 					&							\\ \hline	
\texttt{4-sat-1}        	 &  4         	& 4.31 - 23.3		& 2.5 - 7.8     & 0.0001 - 0.0005   & 3							\\ 
\texttt{4-sat-2}      		 &  4         	& 4.31 - 23.3		& 2.5 - 7.8     & 0.0001 - 0.0005	& 3							\\ 
\texttt{multi-sat-1}    	 &  4        	& 65.2 - 138.4		& 1.1 - 3.0     & 0.0075 - 0.02		& 3							\\ 
\texttt{multi-sat-2}    	 &  5        	& 46.9 - 74.6		& 1.1 - 2.4     & 0.011 - 0.029 	& 4							\\ 
\texttt{multi-sat-3}    	 &  13        	& 2.99 - 29.6	    & 0.67 - 2.5    & 0.004 - 0.11		& 12						\\ 
\texttt{multi-sat-uni-1}	 &  5        	& 6.35 - 46.0    	& 0.83 - 3.1   	&  0.0002 - 0.026	& 2							\\ 
\texttt{multi-sat-uni-1-h015}&  4        	& 17.0 - 45.4    	& 2.36 - 6.97  	&  0.01 - 0.095		& 0							\\ 
\texttt{multi-sat-uni-2-h015}&  7        	& 20.8 - 134.0    	& 0.61 - 4.92  	&  0.022 - 0.15		& 4 						\\ 
\texttt{multi-sat-uni-3-h015}&  6        	& 14.8 - 101.0    	& 1.76 - 6.29  	&  0.115 - 0.036	& 3							\\ 
\texttt{multi-sat-uni-4-h015}&  6        	& 26.2 - 146.1    	& 0.91 - 5.3   	&  0.011 - 0.056	& 1						\\ 
\texttt{multi-sat-uni-5-h015}&  6        	& 13.5 - 87.0    	& 1.82 - 4.42  	&  0.012 - 0.12		& 2							\\ \hline

\end{tabular}
\label{tab:sumary}
\end{table*}

\section{Conclusions}
\label{conclusion} 
We have investigated the growth and orbital evolution of satellite embryos in a circum-planetary disc around a giant planet with the characteristics of Jupiter. For this purpose we used dedicated $N$-body simulations including the effects of type I migration, eccentricity and inclination damping for the embryos and gas drag for the satellitesimals. We simulated a massive, low viscosity disc consistent with the studies regarding MRI regions in Jupiter's circum-planetary disc, we also did not consider any infall of material onto the disc. We performed multiple simulations changing the distribution and the masses of the embryos as well as the number and masses of the satellitesimals in the disc.\par
Previous work on the formation of the Galilean satellites used the type II migration regime \citep{Miguel-Ida-2016}. To verify this possibility we simulated the migration of four massive satellites in a circum-planetary disc using two-dimensional hydrodynamic simulations over 150 years. Our results showed no signs of a gap being opened by the satellites, thus the use of type II migration in this problem is hard to justify. In our hydrodynamic simulations we considered $\alpha=5\times10^{-4}$ for the viscosity prescription, in agreement with recent values for purely hydrodynamic turbulence in discs \citep{Stoll2014A&A...572A..77S, Stoll-etal-2017a, Stoll-etal-2017b}.

The $N$-body simulation with fully formed satellites (having the mass of the Galileans) showed that our integration scheme is capable of reproduce the migration of massive satellites from the outer parts of the disc to a configuration that resembles the current architecture of the Galilean satellites. In addition, we showed that, for massive satellites, initially eccentric and inclined orbits do not affect the final results due to the damping in the disc.
The satellite embryos formed in the outer disc would be enriched with icy materials, as observed,
which is in agreement with our simulations that initially have more massive embryos present in the outer parts of the disc.\par

From the simulations with satellite embryos and satellitesimals we found that the final number of formed satellites depends on the initial distribution of embryos in the disc. Embryos initially located at the outer disc are more likely to survive and become satellites, while the inner embryos tend to migrate faster and collide with the central body due to the high gas density in this region. Also, we showed that most of the satellitesimals in the disc are accreted by the inner embryos or depleted during the migration phase of the embryos, in this way the growth of massive satellites from the outer disc is more difficult. An inflow of mass as proposed in the gas-starved disc \citep{Canup-Ward-2002, Canup-Ward-2006, Canup-Ward-2009} might favour the formation of satellites from the outer disc, also the inflow of material could help the outer satellites to become more massive than the ones we found. The gas-starved disc scenario and the possibility of inflow of mass were not addressed in this work since we were interested in study the evolution of satellites in a static, massive disc. However, even though no inflow of material is a general assumption for the MMSN model, hydro-dynamic simulations have shown that this is no completely true and inflow of material from the
ambient protoplanetary disc might be possible. In this way our models could be improved in the future by adding this effect.    \par
When we simulated more massive embryos, we found a configuration with multiple small satellites close to the planet. These satellites are locked in resonant chains that remain stable after the gas disc dissipates while the eccentricity of the satellites increased due to their proximity. Similar behaviour has been seen in simulations for extrasolar close-in Super-Earths systems \citep{Izidoro-etal-2017}, with the difference that after the gas dissipation the planets start to collide with each other. Thus, we suggest that massive satellites as the Galilean ones and close-in Super-Earths might be formed through similar processes.

In our first set of models the disc surface density has a strong radial dependence ($\Sigma_{gas}\propto r^{-3/2}$)
and the aspect ratio of the disc was $h=0.05$. As a consequence the embryos in the outer disc grow slower than the inner ones, the migration speed of them is slow, and in most of the cases these bodies never reach the inner region where the Jovian satellites are located.
On the other hand, When an uniform distribution for the gaseous disc was used, we found that the outer embryos migrated faster and interacted with the inner ones. In this way the final radial distribution of satellites was improved showing better agreement with the observations. Our results indicate that flatter density distributions with a weak radial dependence favour the migration of satellite embryos from the outer disc.
However, the final masses of the satellites remained an issue.
As in the previous cases, the massive satellites tend to reach the inner cavity of the disc and, due to interactions with other satellites, eventually fall into the planet. \citet{Szulagyi-etal-2016} performed detailed hydrodynamic studies about circum-planetary discs around giant planets and showed that the gaseous disc is flatter in the region of the Galilean satellites corroborating our findings.
\par

The temperature distribution in the disc is an important feature to study the evolution of satellites in a circum-planetary disc. In our first models we adopted the aspect ratio to be $h=0.05$, which is a generally good estimate for $h$ in protoplanetary discs. 
However, in our model for Jupiter's circumplanetary disc this value implies a disc with an ice line very close to the central planet (around $4$ $R_J$), which is in disagreement with the values found in the literature. A well accepted position for the ice line in models like ours is at $30$ $R_J$ \citep{Lunine-Stevenson-1982, Miguel-Ida-2016}. As the initial temperature depends on the aspect ratio (Eq. \ref{eq:t01}) we adopted
in a second sequence of models $h=0.15$. Increasing the value of the aspect ratio also has implications on the migration of the embryos, once they are evolving in a thicker disc it is expected that their migration rate decrease. Our simulations showed that this configuration of the disc favours the formation of massive satellites, $M\ge 6\times 10^{-5}$ $M_J$, which is comparable to the masses of Ganymede and Callisto. We highlight the results from model \texttt{multi-sat-uni-1-h015}, where four satellites were formed with masses that accurately match the masses of the Galilean satellites.\par 

Reproducing the formation and evolution of the Galilean satellites is not a simple task, the processes that led to the formation of these bodies are complicated and full of uncertainties. With this paper we expect to combine results of our simulations to add informations about the origin of the embryos that gave birth to this satellite system, also we found that a flatter initial gas distribution and a disc with ice line around $30$ $R_J$ favours the formation of systems like the Jovian system. We suggest that the similarities between the Galilean satellites system and the close-in Super-Earths systems might lead to a unified theory about the formation of these systems.

\section*{Acknowledgements}
We thank the anonymous referee for the valuable comments and suggestions. We thank Andre Izidoro for the discussions concerning our findings, during the participation in the ``Protoplanetary Disks and Planet Formation and Evolution" programme organized by the Munich Institute for Astro-and Particle Physics (MIAPP). RAM thanks financial support from FAPESP (Grants: 2011/08171-3, 2013/24281-9 and 2016/12113-2).

%I thank Professor N. Kameswara Rao for some helpful suggestions,
%Dr H. C. Bhatt for a critical reading of the original version of the
%paper and an anonymous referee for very useful comments that improved
%the presentation of the paper.
%%%%%%%%%%%%%%%%%%%%%%%%%%%%%%%%%%%%%%%%%%%%%%%%%%



%\bibliographystyle{mnras}

\bibliographystyle{mnras}
\bibliography{ref}

\begin{thebibliography}{}
\makeatletter
\relax
\def\mn@urlcharsother{\let\do\@makeother \do\$\do\&\do\#\do\^\do\_\do\%\do\~}
\def\mn@doi{\begingroup\mn@urlcharsother \@ifnextchar [ {\mn@doi@}
  {\mn@doi@[]}}
\def\mn@doi@[#1]#2{\def\@tempa{#1}\ifx\@tempa\@empty \href
  {http://dx.doi.org/#2} {doi:#2}\else \href {http://dx.doi.org/#2} {#1}\fi
  \endgroup}
\def\mn@eprint#1#2{\mn@eprint@#1:#2::\@nil}
\def\mn@eprint@arXiv#1{\href {http://arxiv.org/abs/#1} {{\tt arXiv:#1}}}
\def\mn@eprint@dblp#1{\href {http://dblp.uni-trier.de/rec/bibtex/#1.xml}
  {dblp:#1}}
\def\mn@eprint@#1:#2:#3:#4\@nil{\def\@tempa {#1}\def\@tempb {#2}\def\@tempc
  {#3}\ifx \@tempc \@empty \let \@tempc \@tempb \let \@tempb \@tempa \fi \ifx
  \@tempb \@empty \def\@tempb {arXiv}\fi \@ifundefined
  {mn@eprint@\@tempb}{\@tempb:\@tempc}{\expandafter \expandafter \csname
  mn@eprint@\@tempb\endcsname \expandafter{\@tempc}}}

\bibitem[\protect\citeauthoryear{{Adachi}, {Hayashi}  \& {Nakazawa}}{{Adachi}
  et~al.}{1976}]{Adachi-etal-1976}
{Adachi} I.,  {Hayashi} C.,   {Nakazawa} K.,  1976, \mn@doi [Progress of
  Theoretical Physics] {10.1143/PTP.56.1756}, \href
  {http://adsabs.harvard.edu/abs/1976PThPh..56.1756A} {56, 1756}

\bibitem[\protect\citeauthoryear{{Brasser}, {Duncan}  \& {Levison}}{{Brasser}
  et~al.}{2007}]{Brasser-etal-2007}
{Brasser} R.,  {Duncan} M.~J.,   {Levison} H.~F.,  2007, \mn@doi [Icarus]
  {10.1016/j.icarus.2007.05.003}, \href
  {http://adsabs.harvard.edu/abs/2007Icar..191..413B} {191, 413}

\bibitem[\protect\citeauthoryear{{Canup} \& {Ward}}{{Canup} \&
  {Ward}}{2002}]{Canup-Ward-2002}
{Canup} R.~M.,  {Ward} W.~R.,  2002, \mn@doi [AJ] {10.1086/344684}, \href
  {http://adsabs.harvard.edu/abs/2002AJ....124.3404C} {124, 3404}

\bibitem[\protect\citeauthoryear{{Canup} \& {Ward}}{{Canup} \&
  {Ward}}{2006}]{Canup-Ward-2006}
{Canup} R.~M.,  {Ward} W.~R.,  2006, \mn@doi [Nature] {10.1038/nature04860},
  \href {http://adsabs.harvard.edu/abs/2006Natur.441..834C} {441, 834}

\bibitem[\protect\citeauthoryear{{Canup} \& {Ward}}{{Canup} \&
  {Ward}}{2009}]{Canup-Ward-2009}
{Canup} R.~M.,  {Ward} W.~R.,  2009, Origin of Europa and the Galilean
  Satellites.
p.~59

\bibitem[\protect\citeauthoryear{{Chambers}}{{Chambers}}{1999}]{Chambers-1999}
{Chambers} J.~E.,  1999, \mn@doi [MNRAS] {10.1046/j.1365-8711.1999.02379.x},
  \href {http://adsabs.harvard.edu/abs/1999MNRAS.304..793C} {304, 793}

\bibitem[\protect\citeauthoryear{{Cresswell} \& {Nelson}}{{Cresswell} \&
  {Nelson}}{2008}]{Cresswell-Nelson-2008}
{Cresswell} P.,  {Nelson} R.~P.,  2008, \mn@doi [A\&A]
  {10.1051/0004-6361:20079178}, \href
  {http://adsabs.harvard.edu/abs/2008A%26A...482..677C} {482, 677}

\bibitem[\protect\citeauthoryear{{Estrada}, {Mosqueira}, {Lissauer}, {D'Angelo}
   \& {Cruikshank}}{{Estrada} et~al.}{2009}]{Estrada-etal-2009}
{Estrada} P.~R.,  {Mosqueira} I.,  {Lissauer} J.~J.,  {D'Angelo} G.,
  {Cruikshank} D.~P.,  2009, Formation of Jupiter and Conditions for Accretion
  of the Galilean Satellites.
p.~27

\bibitem[\protect\citeauthoryear{{Fujii}, {Okuzumi}, {Tanigawa}  \&
  {Inutsuka}}{{Fujii} et~al.}{2014}]{Fujii-etal-2014}
{Fujii} Y.~I.,  {Okuzumi} S.,  {Tanigawa} T.,   {Inutsuka} S.-i.,  2014,
  \mn@doi [ApJ] {10.1088/0004-637X/785/2/101}, \href
  {http://adsabs.harvard.edu/abs/2014ApJ...785..101F} {785, 101}

\bibitem[\protect\citeauthoryear{{Fujii}, {Kobayashi}, {Takahashi}  \&
  {Gressel}}{{Fujii} et~al.}{2017}]{Fujii-etal-2017}
{Fujii} Y.~I.,  {Kobayashi} H.,  {Takahashi} S.~Z.,   {Gressel} O.,  2017,
  \mn@doi [AJ] {10.3847/1538-3881/aa647d}, \href
  {http://adsabs.harvard.edu/abs/2017AJ....153..194F} {153, 194}

\bibitem[\protect\citeauthoryear{{Gillon} et~al.,}{{Gillon}
  et~al.}{2016}]{Gillon-etal-2016}
{Gillon} M.,  et~al., 2016, \mn@doi [Nature] {10.1038/nature17448}, \href
  {http://adsabs.harvard.edu/abs/2016Natur.533..221G} {533, 221}

\bibitem[\protect\citeauthoryear{{Gillon} et~al.,}{{Gillon}
  et~al.}{2017}]{Gillon-etal-2017}
{Gillon} M.,  et~al., 2017, \mn@doi [Nature] {10.1038/nature21360}, \href
  {http://adsabs.harvard.edu/abs/2017Natur.542..456G} {542, 456}

\bibitem[\protect\citeauthoryear{{Greenberg}}{{Greenberg}}{1987}]{Greenberg-1987}
{Greenberg} R.,  1987, \mn@doi [Icarus] {10.1016/0019-1035(87)90139-4}, \href
  {http://adsabs.harvard.edu/abs/1987Icar...70..334G} {70, 334}

\bibitem[\protect\citeauthoryear{{Hayashi}}{{Hayashi}}{1981}]{Hayashi-1981}
{Hayashi} C.,  1981, \mn@doi [Progress of Theoretical Physics Supplement]
  {10.1143/PTPS.70.35}, \href
  {http://adsabs.harvard.edu/abs/1981PThPS..70...35H} {70, 35}

\bibitem[\protect\citeauthoryear{{Heller} \& {Pudritz}}{{Heller} \&
  {Pudritz}}{2015a}]{Heller-Pudritz-2015a}
{Heller} R.,  {Pudritz} R.,  2015a, \mn@doi [A\&A]
  {10.1051/0004-6361/201425487}, \href
  {http://adsabs.harvard.edu/abs/2015A%26A...578A..19H} {578, A19}

\bibitem[\protect\citeauthoryear{{Heller} \& {Pudritz}}{{Heller} \&
  {Pudritz}}{2015b}]{Heller-Pudritz-2015b}
{Heller} R.,  {Pudritz} R.,  2015b, \mn@doi [ApJ]
  {10.1088/0004-637X/806/2/181}, \href
  {http://adsabs.harvard.edu/abs/2015ApJ...806..181H} {806, 181}

\bibitem[\protect\citeauthoryear{{Howard} et~al.,}{{Howard}
  et~al.}{2012}]{Howard-etal-2012}
{Howard} A.~W.,  et~al., 2012, \mn@doi [ApJS] {10.1088/0067-0049/201/2/15},
  \href {http://adsabs.harvard.edu/abs/2012ApJS..201...15H} {201, 15}

\bibitem[\protect\citeauthoryear{{Ida} \& {Lin}}{{Ida} \&
  {Lin}}{2004}]{Ida-Lin-2004}
{Ida} S.,  {Lin} D.~N.~C.,  2004, \mn@doi [ApJ] {10.1086/424830}, \href
  {http://adsabs.harvard.edu/abs/2004ApJ...616..567I} {616, 567}

\bibitem[\protect\citeauthoryear{{Izidoro}, {Morbidelli}  \&
  {Raymond}}{{Izidoro} et~al.}{2014}]{Izidoro-etal-2014}
{Izidoro} A.,  {Morbidelli} A.,   {Raymond} S.~N.,  2014, \mn@doi [ApJ]
  {10.1088/0004-637X/794/1/11}, \href
  {http://adsabs.harvard.edu/abs/2014ApJ...794...11I} {794, 11}

\bibitem[\protect\citeauthoryear{{Izidoro}, {Raymond}, {Pierens}, {Morbidelli},
  {Winter}  \& {Nesvorny`}}{{Izidoro} et~al.}{2016}]{Izidoro-etal-2016}
{Izidoro} A.,  {Raymond} S.~N.,  {Pierens} A.,  {Morbidelli} A.,  {Winter}
  O.~C.,   {Nesvorny`} D.,  2016, \mn@doi [ApJ] {10.3847/1538-4357/833/1/40},
  \href {http://adsabs.harvard.edu/abs/2016ApJ...833...40I} {833, 40}

\bibitem[\protect\citeauthoryear{{Izidoro}, {Ogihara}, {Raymond}, {Morbidelli},
  {Pierens}, {Bitsch}, {Cossou}  \& {Hersant}}{{Izidoro}
  et~al.}{2017}]{Izidoro-etal-2017}
{Izidoro} A.,  {Ogihara} M.,  {Raymond} S.~N.,  {Morbidelli} A.,  {Pierens} A.,
   {Bitsch} B.,  {Cossou} C.,   {Hersant} F.,  2017, \mn@doi [MNRAS]
  {10.1093/mnras/stx1232}, \href
  {http://adsabs.harvard.edu/abs/2017MNRAS.470.1750I} {470, 1750}

\bibitem[\protect\citeauthoryear{{Kley}}{{Kley}}{1999}]{kley-1999}
{Kley} W.,  1999, \mn@doi [MNRAS] {10.1046/j.1365-8711.1999.02198.x}, \href
  {http://adsabs.harvard.edu/abs/1999MNRAS.303..696K} {303, 696}

\bibitem[\protect\citeauthoryear{{Kokubo} \& {Ida}}{{Kokubo} \&
  {Ida}}{2002}]{Kokubo-Ida-2002}
{Kokubo} E.,  {Ida} S.,  2002, \mn@doi [ApJ] {10.1086/344105}, \href
  {http://adsabs.harvard.edu/abs/2002ApJ...581..666K} {581, 666}

\bibitem[\protect\citeauthoryear{{Lubow}, {Seibert}  \& {Artymowicz}}{{Lubow}
  et~al.}{1999}]{lubow}
{Lubow} S.~H.,  {Seibert} M.,   {Artymowicz} P.,  1999, \mn@doi [ApJ]
  {10.1086/308045}, \href {http://adsabs.harvard.edu/abs/1999ApJ...526.1001L}
  {526, 1001}

\bibitem[\protect\citeauthoryear{{Luger} et~al.,}{{Luger}
  et~al.}{2017}]{Luger-etal-2017}
{Luger} R.,  et~al., 2017, \mn@doi [Nature Astronomy]
  {10.1038/s41550-017-0129}, \href
  {http://adsabs.harvard.edu/abs/2017NatAs...1E.129L} {1, 0129}

\bibitem[\protect\citeauthoryear{{Lunine} \& {Stevenson}}{{Lunine} \&
  {Stevenson}}{1982}]{Lunine-Stevenson-1982}
{Lunine} J.~I.,  {Stevenson} D.~J.,  1982, \mn@doi [Icarus]
  {10.1016/0019-1035(82)90166-X}, \href
  {http://adsabs.harvard.edu/abs/1982Icar...52...14L} {52, 14}

\bibitem[\protect\citeauthoryear{{Masset}}{{Masset}}{2000}]{masset}
{Masset} F.,  2000, \mn@doi [A\&A] {10.1051/aas:2000116}, \href
  {http://adsabs.harvard.edu/abs/2000A%26AS..141..165M} {141, 165}

\bibitem[\protect\citeauthoryear{{Mayor} et~al.,}{{Mayor}
  et~al.}{2011}]{Mayor-etal-2011}
{Mayor} M.,  et~al., 2011, preprint, \href
  {http://adsabs.harvard.edu/abs/2011arXiv1109.2497M} {} (\mn@eprint {arXiv}
  {1109.2497})

\bibitem[\protect\citeauthoryear{{Miguel} \& {Ida}}{{Miguel} \&
  {Ida}}{2016}]{Miguel-Ida-2016}
{Miguel} Y.,  {Ida} S.,  2016, \mn@doi [Icarus] {10.1016/j.icarus.2015.10.030},
  \href {http://adsabs.harvard.edu/abs/2016Icar..266....1M} {266, 1}

\bibitem[\protect\citeauthoryear{{Mosqueira} \& {Estrada}}{{Mosqueira} \&
  {Estrada}}{2003a}]{Mosqueira-Estrada-2003a}
{Mosqueira} I.,  {Estrada} P.~R.,  2003a, \mn@doi [Icarus]
  {10.1016/S0019-1035(03)00076-9}, \href
  {http://adsabs.harvard.edu/abs/2003Icar..163..198M} {163, 198}

\bibitem[\protect\citeauthoryear{{Mosqueira} \& {Estrada}}{{Mosqueira} \&
  {Estrada}}{2003b}]{Mosqueira-Estrada-2003b}
{Mosqueira} I.,  {Estrada} P.~R.,  2003b, \mn@doi [Icarus]
  {10.1016/S0019-1035(03)00077-0}, \href
  {http://adsabs.harvard.edu/abs/2003Icar..163..232M} {163, 232}

\bibitem[\protect\citeauthoryear{{Ogihara} \& {Ida}}{{Ogihara} \&
  {Ida}}{2012}]{Ogihara-Ida-2012}
{Ogihara} M.,  {Ida} S.,  2012, \mn@doi [ApJ] {10.1088/0004-637X/753/1/60},
  \href {http://adsabs.harvard.edu/abs/2012ApJ...753...60O} {753, 60}

\bibitem[\protect\citeauthoryear{{Ormel}, {Liu}  \& {Schoonenberg}}{{Ormel}
  et~al.}{2017}]{Ormel-etal-2017}
{Ormel} C.~W.,  {Liu} B.,   {Schoonenberg} D.,  2017, \mn@doi [A\&A]
  {10.1051/0004-6361/201730826}, \href
  {http://adsabs.harvard.edu/abs/2017A%26A...604A...1O} {604, A1}

\bibitem[\protect\citeauthoryear{{Papaloizou} \& {Larwood}}{{Papaloizou} \&
  {Larwood}}{2000}]{Papaloizou-Larwood-2000}
{Papaloizou} J.~C.~B.,  {Larwood} J.~D.,  2000, \mn@doi [MNRAS]
  {10.1046/j.1365-8711.2000.03466.x}, \href
  {http://adsabs.harvard.edu/abs/2000MNRAS.315..823P} {315, 823}

\bibitem[\protect\citeauthoryear{{Peale} \& {Lee}}{{Peale} \&
  {Lee}}{2002}]{Peale-Lee-2002}
{Peale} S.~J.,  {Lee} M.~H.,  2002, \mn@doi [Science]
  {10.1126/science.1076557}, \href
  {http://adsabs.harvard.edu/abs/2002Sci...298..593P} {298, 593}

\bibitem[\protect\citeauthoryear{{Petigura}, {Howard}  \& {Marcy}}{{Petigura}
  et~al.}{2013}]{Petigura-etal-2013}
{Petigura} E.~A.,  {Howard} A.~W.,   {Marcy} G.~W.,  2013, \mn@doi [Proceedings
  of the National Academy of Science] {10.1073/pnas.1319909110}, \href
  {http://adsabs.harvard.edu/abs/2013PNAS..11019273P} {110, 19273}

\bibitem[\protect\citeauthoryear{{Raymond}, {Quinn}  \& {Lunine}}{{Raymond}
  et~al.}{2005}]{Raymond-etal-2005}
{Raymond} S.~N.,  {Quinn} T.,   {Lunine} J.~I.,  2005, \mn@doi [ApJ]
  {10.1086/433179}, \href {http://adsabs.harvard.edu/abs/2005ApJ...632..670R}
  {632, 670}

\bibitem[\protect\citeauthoryear{{Sasaki}, {Stewart}  \& {Ida}}{{Sasaki}
  et~al.}{2010}]{Sasaki-etal-2010}
{Sasaki} T.,  {Stewart} G.~R.,   {Ida} S.,  2010, \mn@doi [ApJ]
  {10.1088/0004-637X/714/2/1052}, \href
  {http://adsabs.harvard.edu/abs/2010ApJ...714.1052S} {714, 1052}

\bibitem[\protect\citeauthoryear{{Shakura} \& {Sunyaev}}{{Shakura} \&
  {Sunyaev}}{1973}]{shakura}
{Shakura} N.~I.,  {Sunyaev} R.~A.,  1973, A\&A, \href
  {http://adsabs.harvard.edu/abs/1973A%26A....24..337S} {24, 337}

\bibitem[\protect\citeauthoryear{{Stoll} \& {Kley}}{{Stoll} \&
  {Kley}}{2014}]{Stoll2014A&A...572A..77S}
{Stoll} M.~H.~R.,  {Kley} W.,  2014, \mn@doi [A\&A]
  {10.1051/0004-6361/201424114}, \href
  {http://adsabs.harvard.edu/abs/2014A%26A...572A..77S} {572, A77}

\bibitem[\protect\citeauthoryear{{Stoll}, {Kley}  \& {Picogna}}{{Stoll}
  et~al.}{2017a}]{Stoll-etal-2017a}
{Stoll} M.~H.~R.,  {Kley} W.,   {Picogna} G.,  2017a, \mn@doi [A\&A]
  {10.1051/0004-6361/201630226}, \href
  {http://adsabs.harvard.edu/abs/2017A%26A...599L...6S} {599, L6}

\bibitem[\protect\citeauthoryear{{Stoll}, {Picogna}  \& {Kley}}{{Stoll}
  et~al.}{2017b}]{Stoll-etal-2017b}
{Stoll} M.~H.~R.,  {Picogna} G.,   {Kley} W.,  2017b, \mn@doi [A\&A]
  {10.1051/0004-6361/201730668}, \href
  {http://adsabs.harvard.edu/abs/2017A%26A...604A..28S} {604, A28}

\bibitem[\protect\citeauthoryear{{Szul{\'a}gyi}, {Masset}, {Lega}, {Crida},
  {Morbidelli}  \& {Guillot}}{{Szul{\'a}gyi} et~al.}{2016}]{Szulagyi-etal-2016}
{Szul{\'a}gyi} J.,  {Masset} F.,  {Lega} E.,  {Crida} A.,  {Morbidelli} A.,
  {Guillot} T.,  2016, \mn@doi [MNRAS] {10.1093/mnras/stw1160}, \href
  {http://adsabs.harvard.edu/abs/2016MNRAS.460.2853S} {460, 2853}

\bibitem[\protect\citeauthoryear{Takata \& Stevenson}{Takata \&
  Stevenson}{1996}]{Takata-Stevenson-1996}
Takata T.,  Stevenson D.~J.,  1996, \mn@doi [Icarus]
  {http://dx.doi.org/10.1006/icar.1996.0167}, 123, 404

\bibitem[\protect\citeauthoryear{{Tanaka} \& {Ward}}{{Tanaka} \&
  {Ward}}{2004}]{Tanaka-Ward-2004}
{Tanaka} H.,  {Ward} W.~R.,  2004, \mn@doi [ApJ] {10.1086/380992}, \href
  {http://adsabs.harvard.edu/abs/2004ApJ...602..388T} {602, 388}

\bibitem[\protect\citeauthoryear{{Tanigawa}, {Ohtsuki}  \&
  {Machida}}{{Tanigawa} et~al.}{2012}]{Tanigawa-etal-2012}
{Tanigawa} T.,  {Ohtsuki} K.,   {Machida} M.~N.,  2012, \mn@doi [ApJ]
  {10.1088/0004-637X/747/1/47}, \href
  {http://adsabs.harvard.edu/abs/2012ApJ...747...47T} {747, 47}

\bibitem[\protect\citeauthoryear{{Turner}, {Lee}  \& {Sano}}{{Turner}
  et~al.}{2014}]{Turner-etal-2014}
{Turner} N.~J.,  {Lee} M.~H.,   {Sano} T.,  2014, \mn@doi [ApJ]
  {10.1088/0004-637X/783/1/14}, \href
  {http://adsabs.harvard.edu/abs/2014ApJ...783...14T} {783, 14}

\bibitem[\protect\citeauthoryear{{Weidenschilling}}{{Weidenschilling}}{1977}]{Weidenschilling-1977}
{Weidenschilling} S.~J.,  1977, \mn@doi [MNRAS] {10.1093/mnras/180.1.57}, \href
  {http://adsabs.harvard.edu/abs/1977MNRAS.180...57W} {180, 57}

\bibitem[\protect\citeauthoryear{{Yoder}}{{Yoder}}{1979}]{Yoder-1979}
{Yoder} C.~F.,  1979, \mn@doi [Nature] {10.1038/279767a0}, \href
  {http://adsabs.harvard.edu/abs/1979Natur.279..767Y} {279, 767}

\bibitem[\protect\citeauthoryear{{Yoder} \& {Peale}}{{Yoder} \&
  {Peale}}{1981}]{Yoder-Peale-1981}
{Yoder} C.~F.,  {Peale} S.~J.,  1981, \mn@doi [Icarus]
  {10.1016/0019-1035(81)90088-9}, \href
  {http://adsabs.harvard.edu/abs/1981Icar...47....1Y} {47, 1}

\makeatother
\end{thebibliography}

%%%%%%%%%%%%%%%%%%%%%%%%%%%%%%%%%%%%%%%%%%%%%%%%%%

% Don't change these lines
\bsp	% typesetting comment
\label{lastpage}
\end{document}